\documentclass[12pt,a4paper]{article}
\usepackage{amsmath,amssymb,bm,ascmac,bbm,mathtools}
\usepackage[dvipdfmx, usenames]{color}
\usepackage[dvipdfmx]{graphicx}
\usepackage{xcolor}
\usepackage{here}
\usepackage{authblk}
\usepackage{lscape}
\newcommand{\mcal}{\mathcal}
\newcommand{\ep}{\epsilon}
\newcommand{\mbb}{\mathbb}

\begin{document}
\title{The fate of non-supersymmetric Gross-Neveu-Yukawa fixed point in two dimensions}
\author[1]{Yu Nakayama}
\author[2]{Ken Kikuchi}
\affil[1]{Department of Physics, Rikkyo University}
\affil[2]{Yau Mathematical Sciences Center, Tsinghua University}
\date{}

\maketitle

\begin{abstract}
We investigate the fate of the non-supersymmetric Gross-Neveu-Yukawa fixed point found by Fei et al in $4-\epsilon$ dimensions with a two-component Majorana fermion continued to two dimensions. Assuming that it is a fermionic minimal model which possesses a chiral $\mathbb{Z}_2$ symmetry (in addition to fermion number parity) and just two relevant singlet operators, we can zero in on four candidates. Assuming further that the least relevant deformation leads to the supersymmetric Gross-Neveu-Yukawa fixed point (i.e. fermionic tricritical Ising model),
we can rule out two of them by matching the spin contents of the preserved topological defect lines. The final candidates are the fermionic $(11,4)$ minimal model if it is non-unitary, and the fermionic $(E_{6}, A_{10})$ minimal model if it is unitary. If we further use a constraint from the double braiding relation proposed by one of the authors,
the former scenario is preferable.

\end{abstract}

\newpage

\section{Introduction}
Minimal models in two-dimensional conformal field theories have been milestones in our understanding of critical phenomena, yet given a concrete Hamiltonian or Lagrangian, it is a non-trivial task to identify which minimal model it corresponds to. In particular, the Landau-Ginzburg descriptions are only known in A-type modular invariant unitary minimal models \cite{Zamolodchikov:1986db} as well as a few examples in D-series \cite{Cardy:1986ie,Fateev:1987vh} (i.e. critical three-state Potts model) and non-unitary theories \cite{Cardy:1985yy} (i.e. Lee-Yang singularities). See also \cite{Li:1988pj,Amoruso,Zambelli:2016cbw,Lencses:2022ira,Klebanov:2022syt} for further attempts.

The difficulty to identify the corresponding minimal model lies in that all the minimal models have essentially the same global symmetries (i.e. $\mathbb{Z}_2$, $\mathbb{Z}_3$ or $\mathbb{Z}_2 \times \mathbb{Z}_2$), and the conventional classification thus resorts to the structure of the operator product expansion and the heuristic renormalization group flow argument. Since they are typically strongly coupled, however, a naive expectation from the Lagrangian description may fail. Under certain limited conditions, the renormalization group flow becomes integrable and the identification of the models becomes easier.

The game changed when we realized that there are more constraints on the renormalization group flow from generalized symmetries \cite{V88,MS1,MS2,GKSW14,BT17,Chang:2018iay}. The idea is to  study more general topological defect lines than the ones that generate group-like symmetries. In two dimensions, we know a lot about these general topological defect lines unlike in higher dimensions.
The spin contents of topological defect lines, for example, are non-trivial renormalization group invariants. There are other constraints on double braidings, global dimensions, scaling dimensions, and so on (in unitary theories).

Equipped with this new machinery, in this paper, we would like to address the fate of the non-supersymmetric Gross-Neveu-Yukawa fixed point in two dimensions that was originally found by Fei et al \cite{Fei:2016sgs} in $4-\epsilon$ dimensions with a two-component Majorana fermion. The essential ingredients of our analysis are the topological defect lines and the associated constraints on the renormalization group flow. We argue that assumptions (1) the fixed point is a (non-unitary) fermionic minimal model, (2) it has a chiral $\mathbb{Z}_2$  symmetry and two relevant singlet operators, and (3) the least relevant deformation leads to the supersymmetric Ising model (i.e. fermionic tricritical Ising model) predict that it is the fermionic  $(11,4)$ minimal model if it is non-unitary, and the fermionic $(E_6, A_{10})$ minimal model if it is unitary. If we further use the constraint from the double braiding relation proposed by one of the authors,
the former scenario is preferable.

The rest of the paper is organized as follows. In section 2, we introduce the Gross-Neveu-Yukawa model with a two-component Majorana fermion and study the non-supersymmetric fixed point. In section 3, we review some properties of topological defect lines and constraints on the renormalization group flow. In section 4, we study the fermionic minimal models to identify the non-supersymmetric fixed points. In section 5, we examine each candidate more in detail. In section 6, we give further discussions and conclude. We have several appendices. In Appendix A, we summarize the convention of the minimal model. In Appendix B, we list the properties of topological defect lines in related models. In Appendix \ref{Z2}, we study   constraints on correlation functions from the action of topological line defects generating anomalous symmetry.

\section{Gross-Neveu-Yukawa model and fixed points}
Let us consider the simplest interacting quantum field theory constructed out of one real scalar  $\phi$ and a two-component Majorana fermion $\psi$. The action is given by
\begin{align}
S = \int d^dx \left(-\frac{1}{2} \partial^\mu \phi \partial_\mu \phi + i\bar{\psi} \gamma_\mu \partial^\mu \psi - g_1 \phi \bar{\psi} \psi - g_2 \phi^4 \right) \ . \label{GNYaction}
\end{align}
In two and three dimensions, the action possesses the (chiral or time-reversal) $\mathbb{Z}_2$ symmetry $\phi \to - \phi$ and $\bar{\psi} \psi \to -\bar{\psi}\psi$. Note that even in fractional dimensions, as long as we use the renormalization group equations based on the minimal subtraction scheme in four dimensions (by further continuing the number of fermions to be $\frac{1}{2}$ because the smallest fermionic degrees of freedom in four dimensions is four rather than two), the   $\mathbb{Z}_2$  symmetry remains in perturbative computations of the renormalization group beta functions. We regard it as a part of the non-perturbative defining property of the Gross-Neveu-Yukawa model with a two-component Majorana fermion.

In \cite{Fei:2016sgs}, they performed a perturbative search of renormalization group fixed points of this model near four dimensions by $\epsilon$ expansions. In addition to the Gaussian fixed point ($g_1=g_2 =0)$ and decoupled Wilson-Fisher fixed point ($g_1$ =0, $g_2\neq 0)$, they found a ``supersymmetric fixed point," which is identified as the supersymmetric Ising model in three dimensions and two dimensions (i.e. fermionic tricritical Ising model or fermionic (5,4) minimal model). What is surprising more is that they also found a ``non-supersymmetric fixed point" whose identification in three dimensions and two dimensions has been elusive. The goal of this paper is to identify the corresponding conformal field theory in two dimensions. 

As observed in \cite{Fei:2016sgs}, in perturbation theory, the non-supersymmetric fixed point has the potential unbounded below and it signals non-unitarity or instability. By continuing the dimension, it may remain non-unitary or may become unitary. We, therefore, do not assume the unitarity of the fixed point in this paper. Rather we make the following three assumptions (i.e. minimality, symmetry, and renormalization group flow) to identify the fate of the fixed point in two dimensions.

The first assumption is the minimality. We assume that the theory is described by a (non-unitary) fermionic minimal model.\footnote{Here we do assume that the fixed point is conformal rather than merely scale invariant. In unitary fixed point, the theorem by Zamolodchikov and Polchinski guarantees it (see e.g. \cite{Nakayama:2013is} for a review). In non-unitary cases, there is no such strong theorem, but the fixed point in $\epsilon$ expansions is conformal, and it is natural to assume that it continues to be the case in two dimensions.} Since we know a complete list of (non-unitary) fermionic minimal models,\footnote{Ferminoic minimal models and bosonic minimal models have the same central charge but have different operator contents. The former includes operators with half-integer spins while the latter does not. Accordingly, the partition functions are different. We will review the construction and the relations between the two in section 4.} this assumption makes our search exhaustive. Note, however, we do not know if this assumption is strictly correct. What we do know is that the central charge of the ultraviolet theory is $c=\frac{3}{2}$ (one free scalar and one Majorana fermion) and the $c$-theorem would suggest that the fixed point has a smaller (effective) central charge. From the action and the naive dimensional analysis, we may argue that the non-supersymmetric fixed point can be obtained by a relevant deformation of the tricritical supersymmetric Ising model with $c=1$ (or the fermionic Ashkin-Teller model), whose action would be 
\begin{align}
S = \int d^dx \left(-\frac{1}{2} \partial^\mu \phi \partial_\mu \phi + i\bar{\psi} \gamma_\mu \partial^\mu \psi - g_1 \phi^2 \bar{\psi} \psi - g_2 \phi^6 \right) \ .
 \label{AT} \end{align}
Then we may argue that the central charge of the non-supersymmetric fixed point we are looking for is less than $1$, making the assumption of minimality more plausible. If the fixed point remains non-unitary, the assumption of minimality could be obscure, but we stick to this assumption (as long as we obtain sensible candidates).

The second assumption is symmetry. In two dimensions, the Hilbert space of fermionic conformal field theories is divided into the Neveu-Schwartz sector and the Ramond sector. The symmetry of the action corresponds to the symmetry of operator product expansions in the Neveu-Schwartz sector. We assume that the chiral $\mathbb{Z}_2$  symmetry of the  Gross-Neveu-Yukawa model remains a symmetry of the fermionic minimal model in the Neveu-Schwartz sector.  Note that the chiral $\mathbb{Z}_2$  symmetry is anomalous and cannot be spontaneously broken unless there is a non-trivial topological field theory that realizes the anomalous symmetry. There exists such a theory indeed: negative mass $\phi^2$ deformation of \eqref{GNYaction} gives a spontaneously broken phase of the anomalous chiral $\mathbb{Z}_2$ symmetry. Because of the existence of such a theory, we cannot rule out the possibility that the non-supersymmetric fixed point may break the chiral $\mathbb{Z}_2$ symmetry, but based on the continuity assumption, we do assume that the fixed point shows the unbroken chiral $\mathbb{Z}_2$ symmetry.

The third assumption is the property of the renormalization group flow. The perturbative fixed point found in $\epsilon$ expansions has two relevant deformations preserving the chiral $\mathbb{Z}_2$  symmetry. We assume that the number of relevant deformations preserving the chiral $\mathbb{Z}_2$  symmetry does not change down to $d=2$. This is believed to be a generic feature of the renormalization group flow with a continuous parameter by regarding the renormalization group flow as a dynamical system \cite{Gukov:2016tnp}. We further assume that there exists a renormalization group flow from the ``non-supersymmetric fixed point" to the ``supersymmetric Ising model" as suggested in $\epsilon$ expansions.

In section 4, we will show that these assumptions dictate that  the candidate of the non-supersymmetric fixed point is the fermionic $(11,4)$ minimal model if it is non-unitary or the fermionic $(E_{6}, A_{10})$ minimal model if it is unitary.  For this purpose, matching (non-)invariants associated with the topological defect lines recently proposed by one of the authors will play a central role, and we will review it in the next section.

\section{Topological defect lines and constraints on renormalization group flow}
We review some properties of topological defect lines and the constraint on renormalization group flow in two dimensions. We are eventually interested in the constraint in fermionic minimal models, but we begin with the bosonic minimal models for illustration. Most of the discussions will also apply to generic rational conformal field theories, but for concreteness, our discussions focus on Virasoro minimal models.

In (bosonic) A-series minimal models, whose partition functions are diagonal modular invariant, a consistent set of topological defect lines is obtained by Verlinde lines. The Verlinde lines $\hat{L}_i$ are labeled by the same label for the primary states $|j\rangle$ on the cylinder and the action is given by
\begin{align}
\hat{L}_i |j \rangle = \frac{S_{ij}}{S_{0j}} |j \rangle \ ,  
\end{align}
where $S_{ij}$ are modular S-matrix of the Virasoro character and $0$ denotes the identity. The consistency of the Verlinde lines is guaranteed by the Verlinde formula:
\begin{align}
N_{ij}^{\ \ k} = \sum_{l} \frac{S_{il} S_{jl} S_{lk}}{S_{0l}}  \ ,
\end{align}
where the fusion coefficients $N_{ij}^{\ \ k}$ are those for the primary operators, but also for the topological defect lines:
\begin{align}
\hat{L}_i \times \hat{L}_j = \sum_k N_{ij}^{\ \ k} \hat{L}_k \ . 
\end{align}
When a topological defect line is invertible, it generates a (global) symmetry of the theory. In A-series minimal models, there is (at most) only one $\mathbb{Z}_2$ symmetry, so most of the Verlinde lines are not invertible.

The action of the topological defect lines on the vacuum state is known as the quantum dimension:
\begin{align}
\hat{L}_i |0 \rangle = d_i |0 \rangle  = \frac{S_{i0}}{S_{00}} |0\rangle \ ,
\end{align}
which is positive for unitary theories. For invertible topological defect lines, we expect that it is unity.\footnote{It, however, turns out that in non-unitary minimal models, the invertible topological defect line that would generate the $\mathbb{Z}_2$ symmetry could have $d=-1$. The consistency with the symmetry of the operator product expansion is discussed in Appendix \ref{Z2}. }

When the action of a topological defect line on a primary operator $\phi_i$ is the same as that on the identity operator, the topological defect line commutes with the primary operator:
\begin{align}
\frac{S_{i0}}{S_{00}} = \frac{S_{ij}}{S_{0j}} \ .
\end{align}
In particular, if the theory is deformed by $\phi_i$, then the topological defect lines that commute with $\phi_i$ must survive under the renormalization group flow.

In A-series (bosonic) minimal models, the defect Hilbert space partition function can be computed by using the action of the defect operators on primary operators as well as the Verlinde formula:
\begin{align}
Z_{L_k} &= \sum_i \hat{L}_{k} (S \chi_i) (S \bar{\chi}_i) \cr
&=
\sum_{i,j} N_{ki}^j \chi^i \bar{\chi}^j \ . 
\end{align}
Note that spin $h_i-\bar{h}_j$ can be non-half-integer. 

When the topological defect line $L_k$ survives under the renormalization group flow, the $F$-symbols are preserved. 
It is regarded as a generalization of the 't Hooft anomaly matching \cite{Chang:2018iay}. Mathematically, the matching of $F$-symbols follows from the fact that the functor from 
 the surviving symmetry category to the infrared symmetry category is monoidal \cite{KKARG}. The $F$-symbols fix spin contents of the defect Hilbert space, so in \cite{emergentSUSY} it was realized that the spin contents associated to surviving topological defect lines should  be invariant under the renormalization group flow. More precisely, heavy operators in the defect Hilbert space may be lifted along the renormalization group flow, and the spin content of a surviving topological defect line in infrared must be a subset of that in ultraviolet.

In the case of conformal field theories where the partition functions are not diagonal modular invariant, there is no (known) systematic construction of topological defect lines in the sense that there is no explicit formula e.g. by the modular S-matrix. The fermionic minimal models we are interested in are such examples. Still, we can proceed as follows. We begin with the torus partition function
\begin{align}
Z = \sum_{i,j} n_{ij} \chi_i \bar{\chi}_j , 
\end{align}
where $\chi_i$ and $\bar{\chi}_i$ are (left/right) Virasoro characters, and $n_{ij}$ are integer multiplicities. We then postulate the action of the topological defect operator $\hat{L}_k$
\begin{align}
Z_{L_k} = \sum_{i,j} L^k_{ij} n_{ij}\chi_i \bar{\chi}_j \ . 
\end{align}
We now perform the S-transformation and check if the (modified) Cardy condition is satisfied by our choice of $L_{ij}^k$. Once we obtain two non-trivial solutions of the (modified) Cardy condition, we fuse them (i.e. consider $L^k_{ij} \times L^{k'}_{ij}$) and then study the consistency with the (modified) Cardy condition until we close the fusion algebra of topological defect lines.

The Cardy condition says that after the S-transformation, the partition function with the topological  defect line inserted should be expanded by the (Virasoro) character with the integer (or semi-positive integer if unitarity is assumed) coefficient. The modified Cardy condition says that it is an integer multiple of $2^{\frac{1}{2}}$. The modified Cardy condition is allowed because of the possibility of a Majorana zero mode living on the defect Hilbert space. The topological defect line is called m-type when it satisfies the (ordinary) Cardy condition while it is called q-type when it satisfies the modified Cardy condition.

Once we obtain the partition function with the topological defect line, one can define the spin content of the defect Hilbert space from reading  $h -\bar{h}$. Similarly to the diagonal modular invariant cases, we claim that the spin content of the preserved topological  defect line must be matched under the renormalization group flow. More precisely, the spin content of the surviving topological defect line of the infrared theory must be contained in that of the ultraviolet theory. It is regarded as a generalization of the 't Hooft anomaly matching.

In this paper, the matching of the spin content of the topological defect line will play the most significant role. There are other renormalization group (non-)invariants associated with the topological defect lines proposed in \cite{KK21,KKII,KKfree} such as the double braiding relations, global dimensions, scaling dimensions, and so on. We will mention them separately when necessary.

\section{Identifying the fixed point in minimal models}
A minimal model in two-dimensional conformal field theory is characterized by two co-prime integers $(p,p')$.\footnote{{We will mostly follow the convention used in the yellow book \cite{DiFrancesco:1997nk}. The only exception is the name of the E-type modular invariant minimal models. They used the convention that $(E_6,A_{p-1})$ minimal models correspond to E-type modular invariant $(p,12)$ minimal models. We, however, opt to use the convention that $(E_6,A_{p'-1})$ minimal models correspond to E-type modular invariant $(12,p')$ minimal models. This change of names assures that $p'<p$ in the models we are interested in.}} The central charge is given by
\begin{align}
c = \bar{c} = 1 - 6\frac{(p-p')}{p p'} \ ,
\end{align}
which is a consequence of the reducibility of the representation of the Virasoro algebra (i.e. minimality assumption). Unitarity requires that $p'=p-1$ or $p=p'-1$. Without loss of generality, we will assume $p'<p$. The unitary minimal models are often characterized by one integer $p'=p-1=m$.

Primary operators of minimal models are characterized by two positive integers $(r,s)$ (as well as their left-moving counterpart), and these two integers determine the conformal weights by the formula:
\begin{align}
h_{rs} &= \frac{(pr-p's)^2 - (p-p')^2}{4pp'}  \ ,
\end{align}
where $r<p'$ and $s<p$. The list of primary operators is degenerate due to the identification $(r,s) = (p'-r,p-s)$. 

The modular S-matrix for the Virasoro character in minimal models is explicitly given by
\begin{align}
S_{rs,\rho\sigma} = 2\sqrt{\frac{2}{pp'}} (-1)^{1+ s \rho + r \sigma} \sin(\pi \frac{p}{p'} r \rho)\sin(\pi \frac{p'}{p} s\sigma) \ ,
\end{align}
which is real and symmetric.

In a classic paper \cite{Zamolodchikov:1986gt}, Zamolodchikov showed that along the renormalization group flow that connects two unitary conformal field theories, the central charge decrease $c^{\mathrm{UV}}> c^{\mathrm{IR}}$. 
In non-unitary conformal field theories, it may be more useful to define the effective central charge
\begin{align}
c_{\mathrm{eff}} = \bar{c}_{\mathrm{eff}} = c - 24h_{\mathrm{min}}
= 1 - \frac{6}{p p'} \ ,
\end{align}
where $h_{\mathrm{min}}$ is the lowest conformal dimension of primary operators in the theory.
It was then proposed that the effective central charge is non-increasing $c^{\mathrm{UV}}_{\mathrm{eff}} \ge c^{\mathrm{IR}}_{\mathrm{eff}}$ along the renormalization group flow that connects two $\mathcal{PT}$ invariant conformal field theories (without unitarity) \cite{Castro-Alvaredo:2017udm}.

Our target is fermionic minimal models. As discussed in \cite{Petkova:1988cy,Hsieh:2020uwb}, fermionic minimal models can be constructed when the bosonic minimal model has a non-anomalous $\mathbb{Z}_2$ symmetry. We begin with the A-type  (or D-type) modular invariant bosonic minimal models.\footnote{Since we are only interested in the Neveu-Schwartz sector, there is no distinction between the fermionic A-series minimal model and the fermionic D-series minimal model.} The fusion rule of the (chiral half of the) primary operators
\begin{align}
\phi_{(r_1,s_1)} \otimes \phi_{(r_2,s_2)} = \sum_{r_3\equiv|r_1-r_2|-1}^{\text{min}(r_1+r_2,2p'-r_1-r_2)-1}\sum_{s_3\equiv|s_1-s_2|-1}^{\text{min}(s_1+s_2,2p-s_1-s_2)-1} \phi_{(r_3,s_3)} \ ,
\end{align}
where the summation is over positive integers modulo two, has the following $\mathbb{Z}_2$ symmetry \cite{Ruelle:1998zu}
\begin{align}
r-1 \ \text{mod} \ 2 \ \ (\text{when $p'$ is even}) \cr
s-1 \ \text{mod} \ 2  \ \ (\text{when $p$ is even}) \cr
r+s-2 \ \text {mod} \ 2  \ \ (\text{when $p'$ and $p$ are both odd})  .
\end{align}
It is not $\mathbb{Z}_2 \times \mathbb{Z}_2$ due to the operator identification $(r,s) = (p'-r,p-s)$.
It turns out the last case is anomalous (i.e. it cannot be gauged), so the fermionic minimal models exist only when either $p'$ or $p$ is even.\footnote{The exception is when $p'$ or $p$ is two, where there is no $\mathbb{Z}_2 $ symmetry.} 

To complete the list, let us add the E-series fermionic minimal models. Within unitary minimal models, they are at $m=11,12$ \cite{Kulp:2020iet}. More generally, allowing for non-unitarity, they exist for $(E_6,A_{p'-1})$ minimal models (where $p'$ must be co-prime with 12). The other E-series minimal models do not possess the $\mathbb{Z}_2$ symmetry, and the fermionic counterparts do not exist.

Let us now consider the chiral $\mathbb{Z}_2$  symmetry of the fermionic minimal models. Generic fermionic minimal models do not possess $\mathbb{Z}_2$ symmetries except for the (non-chiral) fermionic parity $(-1)^F$ and the Neveu-Schwartz/Ramond parity. The chiral 
 $\mathbb{Z}_2$ symmetry acts on the Neveu-Schwartz sector non-trivially and it must be inherent from the bosonic minimal models as a selection rule of the operator product expansions. In other words, the corresponding bosonic minimal model must have a hidden $\mathbb{Z}_2$ symmetry inside the singlet sector of the $\mathbb{Z}_2$ symmetry that we use to construct a fermionic minimal model.  We claim that the chiral $\mathbb{Z}_2$ symmetry exists only in $(p,4)$ fermionic minimal models and exceptional $(E_6,A_{p'-1})$ fermionic minimal models.
 
 The chiral or hidden $\mathbb{Z}_2$ symmetry in the Ising model and the tricritical Ising model is well-known: it is unexpected from the Landau-Ginzburg viewpoint but it is a manifest symmetry of the action in the fermionic description. In the correspondence between the Ising model and the free Majorana fermion, the flip of fermion bilinear $\bar{\psi}\psi \to -\bar{\psi}{\psi}$ corresponds to $\epsilon \to -\epsilon$, and we recall that the Ising model has a miraculous operator product expansion 
 \begin{align}
\epsilon \times \epsilon = 1 
 \end{align}
 and so does the tricritical Ising model.\footnote{In the bosonic description, the symmetry is associated with the Kramers–Wannier duality of the two-dimensional lattice model, which states that the low-temperature phase and the high-temperature phase can be related. \label{KWfoot}}
 Higher A-series or D-series unitary minimal models $m>4$ do not possess such symmetries in the operator product expansions. 
 
In the bosonic description, the reason why we have the extra  $\mathbb{Z}_2$ symmetry in the $\mathbb{Z}_2$ invariant subsector of the $(p,4)$ minimal models is due to the fact that the Kac table of the primary operators consists of only three rows, and the operator product expansions of $(1,s)$ operator close, where the chiral $\mathbb{Z}_2$ symmetry acts as even/odd parity of $s$. 

The chiral $\mathbb{Z}_2$ symmetry of the $(E_6,A_{p'-1})$ fermionic minimal model has a different origin. The Neveu-Schwartz sector partition function has the form
\begin{align}
Z_{\mathrm{NS}} = &\sum_{1\le r \le \frac{p'-1}{2}} |\chi_{r,1} + \chi_{r,7}|^2 + |\chi_{r,5} + \chi_{r,11}|^2 \cr 
=& \sum_{1\le r \le \frac{p'-1}{2}} |C_{r,1}|^2 + |C_{r,5}|^2 
\end{align}
and the $\mathbb{Z}_2$ acts non-trivially on $\chi_{\text{even},1}, \chi_{\text{even},7}, \chi_{\text{odd},5}, \chi_{\text{odd},11}$, and trivially otherwise. One can directly check that it is the symmetry of the singlet sector of the bosonic $(E_6,A_{p'-1})$ minimal models from the explicit form of the fusion rule.

As we have already mentioned in footnote \ref{KWfoot}, the existence of the chiral $\mathbb{Z}_2$ symmetry in the fermionic minimal models should be related to the existence of the Kramers-Wannier type duality in the bosonic counterparts. The Kramers-Wannier duality is associated with the topological defect line $N$, which satisfies the fusion rule $N \times N = 1 + \eta$, where $\eta$ is an invertible topological line defect that generates the (non-chiral) $\mathbb{Z}_2$ symmetry. In Appendix B, we will explicitly show the existence of the Kramers-Wannier duality line $N$ in the bosonic counterpart of our candidates.

We now discuss the constraint from the renormalization group flow. Among $(4,p)$ fermionic minimal models and $(E_6,A_{p'-1})$ fermionic minimal models, $(9,4)$, $(11,4)$, $(E_6,A_6)$, $(E_6,A_{10})$ fermionic minimal models have two relevant operators that are singlet under the chiral $\mathbb{Z}_2$ symmetry. Among the four, only the $(E_6,A_{10})$ fermionic minimal model is unitary, and the other three models are non-unitary. 

All of the four candidates have a larger effective central charge than the supersymmetric Ising model $c=\frac{7}{10}$, so they may flow to the supersymmetric Ising model by a relevant deformation that is invariant under the chiral $\mathbb{Z}_2$ symmetry. In this renormalization group flow, the fixed point is approached by the $T\bar{T}$ operator (i.e . dimension four energy-momentum tensor bilinear), which is the least relevant singlet operator that is not a total derivative. It should be contrasted with the integrable flow from the fermionic $(6,5)$ minimal models, where it is approached by the $|\epsilon_{\frac{3}{2}}|^2$ operator, which breaks the chiral $\mathbb{Z}_2$ symmetry explicitly.\footnote{We recall that the renormalization group flow from the (fermionic) $(5,4)$ minimal model to the (fermionic) $(4,3)$ minimal model is approached by the $T\bar{T}$ operator, which is the least relevant singlet operator that is not a total derivative. See e.g. \cite{Zamolodchikov:1991vx}. The $T\bar{T}$ deformation has been actively studied in recent literature stemming from \cite{Smirnov:2016lqw}.}

To further give a refined constraint, we will study the details of the renormalization group flow by matching renormalization group (non-)invariants associated with the topological defect lines proposed in \cite{KK21,emergentSUSY,KKII,KKfree}. Our strategy is to find topological defect lines that commute with the deformation and see whether they match the ones in the supersymmetric Ising model. The main focus is the spin content of the preserved topological defect lines, and their matching is regarded as a generalization of the 't Hooft anomaly matching.

\section{Four candidates}

In this section, we would like to study the four candidates one by one. For this purpose, we first review the fermionic $(5,4)$ minimal model (i.e. supersymmetric Ising model or fermionic tricritical Ising model) that should be in the same universality class as the Gross-Neveu-Yukawa fixed point with a two-component Majorana fermion in two dimensions. It has the central charge $c=\frac{7}{10}$. 

\underline{Primary operators and Verlinde lines}

We list the primary operators in the Neveu-Schwartz sector.
In the spin-zero sector, we may borrow the identification of the Verlinde lines with the primary operators in the bosonic A-series minimal models, which is shown in the last entry of the table below.
\begin{align*}
    \phi_{(1,1)}\bar{\phi}_{(1,1)}=\phi_{(3,4)}\bar{\phi}_{(3,4)} &\leftrightarrow id_{0,0} \leftrightarrow1,\\
    \phi_{(1,2)}\bar{\phi}_{(1,2)}=\phi_{(3,3)}\bar{\phi}_{(3,3)} &\leftrightarrow|\epsilon_{\frac{1}{10}}|^2 \leftrightarrow(-1)^FW, \\  
    \phi_{(1,3)}\bar{\phi}_{(1,3)}=\phi_{(3,2)}\bar{\phi}_{(3,2)} &\leftrightarrow|\epsilon_{\frac{3}{5}}|^2\leftrightarrow W, \\
    \phi_{(1,4)}\bar{\phi}_{(1,4)}=\phi_{(3,1)}\bar{\phi}_{(3,1)} &\leftrightarrow|\epsilon_{\frac{3}{2}}|^2\leftrightarrow(-1)^F, \\
    \phi_{(1,1)}\bar{\phi}_{(1,4)}=\phi_{(3,4)}\bar{\phi}_{(3,1)} &\leftrightarrow \bar{\epsilon}_{\frac{3}{2}},\\
    \phi_{(1,2)}\bar{\phi}_{(1,3)}=\phi_{(3,3)}\bar{\phi}_{(3,2)} &\leftrightarrow\epsilon_{\frac{1}{10}}\bar{\epsilon}_{\frac{3}{5}}, \\  
    \phi_{(1,3)}\bar{\phi}_{(1,2)}=\phi_{(3,2)}\bar{\phi}_{(3,3)} &\leftrightarrow \epsilon_{\frac{3}{5}}\bar{\epsilon}_{\frac{1}{10}} , \\
    \phi_{(1,4)}\bar{\phi}_{(1,1)}=\phi_{(3,1)}\bar{\phi}_{(3,4)} &\leftrightarrow \epsilon_{\frac{3}{2}}, \\
\end{align*}
{The theory has a supersymmetry, which is generated by the fermionic current $\epsilon_{\frac{3}{2}}$ and $\bar{\epsilon}_{\frac{3}{2}}$.}

\underline{Action of topological defect lines on primaries}

{We find 8 topological defect lines (4 m-type and 4 q-type) with the action}
\begin{table}[H]
\begin{center}
\begin{tabular}{c|c|c|c|c||c|c|c|c}
	NS-NS&$id$&$|\ep_{\frac1{10}}|^2$&$|\ep_{\frac35}|^2$&$|\ep_{\frac32}|^2$&$\ep_{\frac32}$&$\bar\ep_{\frac32}$&$\ep_{\frac1{10}}\bar\ep_{\frac35}$&$\ep_{\frac35}\bar\ep_{\frac1{10}}$\\\hline
	$(-1)^F$&$1$&$1$&$1$&$1$&$-1$&$-1$&$-1$&$-1$\\
	$W$&$\zeta$&$-\zeta^{-1}$&$-\zeta^{-1}$&$\zeta$&$\zeta$&$\zeta$&$-\zeta^{-1}$&$-\zeta^{-1}$\\
	$R$&$1$&$-1$&$1$&$-1$&$-1$&$1$&$-1$&$1$
\end{tabular}.
\end{center}
\end{table}
Here $\zeta = \frac{1+\sqrt{5}}{2}$.
The invertible line $R$ generates the chiral $\mathbb{Z}_2$ symmetry. Note that $R$ is not the Verlinde line of the bosonic A-series minimal model. The closest object is the Kramers-Wannier duality line $N$ {(see Appendix B)}. The chiral $\mathbb{Z}_2$ symmetry generated by $R$ does not commute with the supersymmetry generated by $\epsilon_{\frac{3}{2}}$, so it is the R-symmetry.

\underline{Spin contents}

By performing the modular S-transformation, one can compute the spin contents of the topological defect lines. We only show here those that will be matched along the renormalization group flow:
\begin{align*}
    \mcal H_{(-1)^F}:&\quad s\in\{0\},\\
    \mcal H_{R}:&\quad s\in \{-\frac7{16},\frac1{16},\frac9{16},\frac{17}{16}\},\\
    \mcal H_{(-1)^F R}:&\quad s\in \{-\frac{17}{16},-\frac9{16},-\frac1{16},\frac7{16}\}.
\end{align*}

As we have already discussed, we demand that the non-supersymmetric Gross-Neveu-Yukawa fixed point will flow to this supersymmetric fixed point due to the relevant singlet deformation. We will study the consistency of the putative flow in each candidate.

\subsection{Fermionic $(11,4)$ minimal model}

We begin with the fermionic $(11,4)$ minimal model with $c=-\frac{125}{22}$ and $c_{\mathrm{eff}} = \frac{19}{22} \approx 0.86$. We will see that it is a promising (non-unitary) candidate, which passes all the constraints of the renormalization group flow that we are aware of. Indeed, the renormalization group flow from the bosonic $(11,4)$ minimal model to the bosonic $(5,4)$ minimal model induced by the least relevant deformation $\phi_{(1,5)}\bar{\phi}_{(1,5)}$ was studied in \cite{Dorey:2000zb} by using a nonlinear integral equation. Since the perturbation used in the bosonic minimal models is singlet under the chiral $\mathbb{Z}_2$ symmetry, it is highly expected the corresponding flow in the fermionic minimal models should be consistent.

\underline{Primary operator and topological defect lines}

We first  list  primary operators in the Neveu-Schwartz sector and corresponding (Verlinde-like) topological defect lines:
\begin{align*}
    \phi_{(1,1)}\bar\phi_{(1,1)}=\phi_{(3,10)}\bar\phi_{(3,10)}&\leftrightarrow id_{0,0}\leftrightarrow1,\\
    \phi_{(3,1)}\bar\phi_{(3,1)}=\phi_{(1,10)}\bar\phi_{(1,10)}&\leftrightarrow|\ep_{\frac92}|^2\leftrightarrow(-1)^F,\\
    \phi_{(1,2)}\bar\phi_{(1,2)}=\phi_{(3,9)}\bar\phi_{(3,9)}&\leftrightarrow|\ep_{-\frac5{22}}|^2\leftrightarrow\mcal L_{-\frac5{22},-\frac5{22}},\\
    \phi_{(3,2)}\bar\phi_{(3,2)}=\phi_{(1,9)}\bar\phi_{(1,9)}&\leftrightarrow|\ep_{\frac{36}{11}}|^2\leftrightarrow\mcal L_{\frac{36}{11},\frac{36}{11}},\\
    \phi_{(1,3)}\bar\phi_{(1,3)}=\phi_{(3,8)}\bar\phi_{(3,8)}&\leftrightarrow|\ep_{-\frac3{11}}|^2\leftrightarrow\mcal L_{-\frac3{11},-\frac3{11}},\\
    \phi_{(3,3)}\bar\phi_{(3,3)}=\phi_{(1,8)}\bar\phi_{(1,8)}&\leftrightarrow|\ep_{\frac{49}{22}}|^2\leftrightarrow\mcal L_{\frac{49}{22},\frac{49}{22}},\\
    \phi_{(1,4)}\bar\phi_{(1,4)}=\phi_{(3,7)}\bar\phi_{(3,7)}&\leftrightarrow|\ep_{-\frac3{22}}|^2\leftrightarrow\mcal L_{-\frac3{22},-\frac3{22}},\\
    \phi_{(3,4)}\bar\phi_{(3,4)}=\phi_{(1,7)}\bar\phi_{(1,7)}&\leftrightarrow|\ep_{\frac{15}{11}}|^2\leftrightarrow\mcal L_{\frac{15}{11},\frac{15}{11}},\\
    \phi_{(1,5)}\bar\phi_{(1,5)}=\phi_{(3,6)}\bar\phi_{(3,6)}&\leftrightarrow|\ep_{\frac2{11}}|^2\leftrightarrow\mcal L_{\frac2{11},\frac2{11}},\\
    \phi_{(3,5)}\bar\phi_{(3,5)}=\phi_{(1,6)}\bar\phi_{(1,6)}&\leftrightarrow|\ep_{\frac{15}{22}}|^2\leftrightarrow\mcal L_{\frac{15}{22},\frac{15}{22}},\\
    \phi_{(1,1)}\bar\phi_{(3,1)}&\leftrightarrow\bar\ep_{\frac92},\\
    \phi_{(3,1)}\bar\phi_{(1,1)}&\leftrightarrow\ep_{\frac92},\\
    \phi_{(1,2)}\bar\phi_{(3,2)}&\leftrightarrow\ep_{-\frac5{22}}\bar\ep_{\frac{36}{11}},\\
    \phi_{(3,2)}\bar\phi_{(1,2)}&\leftrightarrow\ep_{\frac{36}{11}}\bar\ep_{-\frac5{22}},\\
    \phi_{(1,3)}\bar\phi_{(3,3)}&\leftrightarrow\ep_{-\frac3{11}}\bar\ep_{\frac{49}{22}},\\
    \phi_{(3,3)}\bar\phi_{(1,3)}&\leftrightarrow\ep_{\frac{49}{22}}\bar\ep_{-\frac3{11}},\\
    \phi_{(1,4)}\bar\phi_{(3,4)}&\leftrightarrow\ep_{-\frac3{22}}\bar\ep_{\frac{15}{11}},\\
    \phi_{(3,4)}\bar\phi_{(1,4)}&\leftrightarrow\ep_{\frac{15}{11}}\bar\ep_{-\frac3{22}},\\
    \phi_{(1,5)}\bar\phi_{(3,5)}&\leftrightarrow\ep_{\frac2{11}}\bar\ep_{\frac{15}{22}},\\
    \phi_{(3,5)}\bar\phi_{(1,5)}&\leftrightarrow\ep_{\frac{15}{22}}\bar\ep_{\frac2{11}}.
\end{align*}
Note that it has a fermionic chiral current of spin $\frac{9}{2}$.

\underline{Action of topological defect lines on primaries}\\
Solving the (modified) Cardy condition, we find 20 topological defect lines (10 m-type and 10 q-type) with actions
\begin{table}[H]
\hspace{-58pt}
\scalebox{0.49}{\begin{tabular}{c||c|c|c|c|c|c|c|c|c|c||c|c|c|c|c|c|c|c|c|c}
NS-NS&$id$&$|\ep_{\frac92}|^2$&$|\ep_{-\frac5{22}}|^2$&$|\ep_{\frac{36}{11}}|^2$&$|\ep_{-\frac3{11}}|^2$&$|\ep_{\frac{49}{22}}|^2$&$|\ep_{-\frac3{22}}|^2$&$|\ep_{\frac{15}{11}}|^2$&$|\ep_{\frac2{11}}|^2=|\phi_{(1,5)}|^2$&$|\ep_{\frac{15}{22}}|^2$&$\bar\ep_{\frac92}$&$\ep_{\frac92}$&$\ep_{-\frac5{22}}\bar\ep_{\frac{36}{11}}$&$\ep_{\frac{36}{11}}\bar\ep_{-\frac5{22}}$&$\ep_{-\frac3{11}}\bar\ep_{\frac{49}{22}}$&$\ep_{\frac{49}{22}}\bar\ep_{-\frac3{11}}$&$\ep_{-\frac3{22}}\bar\ep_{\frac{15}{11}}$&$\ep_{\frac{15}{11}}\bar\ep_{-\frac3{22}}$&$\ep_{\frac2{11}}\bar\ep_{\frac{15}{22}}$&$\ep_{\frac{15}{22}}\bar\ep_{\frac2{11}}$\\\hline
$(-1)^F$&$1$&$1$&$1$&$1$&$1$&$1$&$1$&$1$&$1$&$1$&$-1$&$-1$&$-1$&$-1$&$-1$&$-1$&$-1$&$-1$&$-1$&$-1$\\\hline
$\mcal L$&$1$&$-1$&$-1$&$1$&$1$&$-1$&$-1$&$1$&$1$&$-1$&$1$&$-1$&$-1$&$1$&$1$&$-1$&$-1$&$1$&$1$&$-1$\\\hline
$\mcal L_{-\frac5{22},-\frac5{22}}$&$-2\sin\frac{3\pi}{22}$&$-2\sin\frac{3\pi}{22}$&$2\sin\frac{5\pi}{22}$&$2\sin\frac{5\pi}{22}$&$2\cos\frac\pi{11}$&$2\cos\frac\pi{11}$&$2\sin\frac\pi{22}$&$2\sin\frac\pi{22}$&$-\frac{\cos\frac{3\pi}{22}}{\sin\frac{2\pi}{11}}$&$-\frac{\cos\frac{3\pi}{22}}{\sin\frac{2\pi}{11}}$&$-2\sin\frac{3\pi}{22}$&$-2\sin\frac{3\pi}{22}$&$2\sin\frac{5\pi}{22}$&$2\sin\frac{5\pi}{22}$&$2\cos\frac\pi{11}$&$2\cos\frac\pi{11}$&$2\sin\frac\pi{22}$&$2\sin\frac\pi{22}$&$-\frac{\cos\frac{3\pi}{22}}{\sin\frac{2\pi}{11}}$&$-\frac{\cos\frac{3\pi}{22}}{\sin\frac{2\pi}{11}}$\\\hline
$\mcal L_{-\frac3{11},-\frac3{11}}$&$-\frac{\sin\frac\pi{11}}{\cos\frac{3\pi}{22}}$&$-\frac{\sin\frac\pi{11}}{\cos\frac{3\pi}{22}}$&$\frac{\sin\frac{2\pi}{11}}{\cos\frac{5\pi}{22}}$&$\frac{\sin\frac{2\pi}{11}}{\cos\frac{5\pi}{22}}$&$\frac{\cos\frac{5\pi}{22}}{\sin\frac\pi{11}}$&$\frac{\cos\frac{5\pi}{22}}{\sin\frac\pi{11}}$&$1-2\cos\frac\pi{11}$&$1-2\cos\frac\pi{11}$&$\frac{\cos\frac\pi{22}}{\sin\frac{2\pi}{11}}$&$\frac{\cos\frac\pi{22}}{\sin\frac{2\pi}{11}}$&$-\frac{\sin\frac\pi{11}}{\cos\frac{3\pi}{22}}$&$-\frac{\sin\frac\pi{11}}{\cos\frac{3\pi}{22}}$&$\frac{\sin\frac{2\pi}{11}}{\cos\frac{5\pi}{22}}$&$\frac{\sin\frac{2\pi}{11}}{\cos\frac{5\pi}{22}}$&$\frac{\cos\frac{5\pi}{22}}{\sin\frac\pi{11}}$&$\frac{\cos\frac{5\pi}{22}}{\sin\frac\pi{11}}$&$1-2\cos\frac\pi{11}$&$1-2\cos\frac\pi{11}$&$\frac{\cos\frac\pi{22}}{\sin\frac{2\pi}{11}}$&$\frac{\cos\frac\pi{22}}{\sin\frac{2\pi}{11}}$\\\hline
$\mcal L_{-\frac3{22},-\frac3{22}}$&$\frac1{-1+2\cos\frac\pi{11}}$&$\frac1{-1+2\cos\frac\pi{11}}$&$-\frac{\sin\frac\pi{11}}{\cos\frac{5\pi}{22}}$&$-\frac{\sin\frac\pi{11}}{\cos\frac{5\pi}{22}}$&$\frac{\cos\frac{3\pi}{22}}{\sin\frac\pi{11}}$&$\frac{\cos\frac{3\pi}{22}}{\sin\frac\pi{11}}$&$-\frac{\sin\frac{2\pi}{11}}{\cos\frac\pi{22}}$&$-\frac{\sin\frac{2\pi}{11}}{\cos\frac\pi{22}}$&$-\frac{\cos\frac{5\pi}{22}}{\sin\frac{2\pi}{11}}$&$-\frac{\cos\frac{5\pi}{22}}{\sin\frac{2\pi}{11}}$&$\frac1{-1+2\cos\frac\pi{11}}$&$\frac1{-1+2\cos\frac\pi{11}}$&$-\frac{\sin\frac\pi{11}}{\cos\frac{5\pi}{22}}$&$-\frac{\sin\frac\pi{11}}{\cos\frac{5\pi}{22}}$&$\frac{\cos\frac{3\pi}{22}}{\sin\frac\pi{11}}$&$\frac{\cos\frac{3\pi}{22}}{\sin\frac\pi{11}}$&$-\frac{\sin\frac{2\pi}{11}}{\cos\frac\pi{22}}$&$-\frac{\sin\frac{2\pi}{11}}{\cos\frac\pi{22}}$&$-\frac{\cos\frac{5\pi}{22}}{\sin\frac{2\pi}{11}}$&$-\frac{\cos\frac{5\pi}{22}}{\sin\frac{2\pi}{11}}$\\\hline
$\mcal L_{\frac2{11},\frac2{11}}$&$-\frac{\sin\frac{2\pi}{11}}{\cos\frac{3\pi}{22}}$&$-\frac{\sin\frac{2\pi}{11}}{\cos\frac{3\pi}{22}}$&$-\frac1{2\sin\frac{3\pi}{22}}$&$-\frac1{2\sin\frac{3\pi}{22}}$&$\frac1{2\sin\frac\pi{22}}$&$\frac1{2\sin\frac\pi{22}}$&$\frac1{2\sin\frac{5\pi}{22}}$&$\frac1{2\sin\frac{5\pi}{22}}$&$\frac1{2\cos\frac\pi{11}}$&$\frac1{2\cos\frac\pi{11}}$&$-\frac{\sin\frac{2\pi}{11}}{\cos\frac{3\pi}{22}}$&$-\frac{\sin\frac{2\pi}{11}}{\cos\frac{3\pi}{22}}$&$-\frac1{2\sin\frac{3\pi}{22}}$&$-\frac1{2\sin\frac{3\pi}{22}}$&$\frac1{2\sin\frac\pi{22}}$&$\frac1{2\sin\frac\pi{22}}$&$\frac1{2\sin\frac{5\pi}{22}}$&$\frac1{2\sin\frac{5\pi}{22}}$&$\frac1{2\cos\frac\pi{11}}$&$\frac1{2\cos\frac\pi{11}}$
\end{tabular}.}
\end{table}

The $\mathcal{L}$ line, which does not exist in the bosonic theory, generates the chiral $\mathbb{Z}_2$ symmetry.
From the action, we learn relevant operators  $|\ep_{-\frac3{11}}|^2$  and $|\phi_{(1,5)}|^2=|\ep_{\frac2{11}}|^2$  preserve four topological defect lines $\{1,(-1)^F,\mcal L,(-1)^F\mcal L\}$. 

\underline{Spin contents}

Performing the modular S-transformation, we can read the spin contents of the topological defect lines that are preserved under the renormalization group flow.
\begin{align*}
    \mcal H_{(-1)^F}:&\quad s\in\{0\},\\
    \mcal H_{\mcal L}:&\quad s\in\{-\frac{47}{16},-\frac{39}{16},-\frac{31}{16},-\frac{23}{16},-\frac{15}{16},-\frac7{16},\frac1{16},\frac9{16},\frac{17}{16},\frac{25}{16}\}\\
    &~~~~~~=\{-\frac7{16},\frac1{16}\}\text{ mod }1,\\
    \mcal H_{(-1)^F\mcal L}:&\quad s\in\{-\frac{25}{16},-\frac{17}{16},-\frac9{16},-\frac1{16},\frac7{16},\frac{15}{16},\frac{23}{16},\frac{31}{16},\frac{39}{16},\frac{47}{16}\}\\
    &~~~~~~=\{-\frac1{16},\frac7{16}\}\text{ mod }1.
\end{align*}
We see that the spin contents agree with those in the fermionic $(5,4)$ minimal model.

In this scenario, we propose the following identifications of (relevant) primary operators with the Lagrangian theories under the renormalization group flow from the non-supersymmetric fixed point (i.e. fermionic (11,4) minimal mode) to the supersymmetric fixed point (i.e. fermionic (5,4) minimal model).
\begin{align}
\phi &\sim |\epsilon_{-\frac{5}{22}}|^2\to |\epsilon_{\frac{1}{10}}|^2 \cr
\phi^2 & \sim  |\epsilon_{-\frac{3}{11}}|^2\to |\epsilon_{\frac{3}{5}}|^2 , \cr
\phi^3 , \bar{\psi}\psi & \sim   |\epsilon_{-\frac{3}{22}}|^2 , |L_{-1}\epsilon_{-\frac{5}{22}}|^2  \to|\epsilon_{\frac{3}{2}}|^2 , |L_{-1}\epsilon_{\frac{1}{10}}|^2 
\cr
\phi^4, \phi \psi \bar{\psi} &\sim |\epsilon_{\frac{2}{11}}|^2, |\epsilon_{\frac{15}{11}}|^2 \to |L_{-2}|^2, |L_{-2} \epsilon_{\frac{3}{5}}|^2,
\cr
\psi & \sim  \epsilon_\frac{15}{22} \bar{\epsilon}_\frac{2}{11} \to \epsilon_{\frac{3}{5}} \bar{\epsilon}_\frac{1}{10} .
 \end{align}
We see that there is one more relevant $\mathbb{Z}_2$ odd primary operator $|\epsilon_{\frac{15}{22}}|^2$  than in the $(5,4)$ fermionic minimal model, which may be unexpected but it is not prohibited by any renormalization group argument.
 
 Matching the spin contents, we can also identify the preserved topological defect lines:
\[ \begin{array}{ccccc}
\text{fermionic }(11,4):&1&(-1)^F&\mcal L&(-1)^F\mcal L\\
&\downarrow&\downarrow&\downarrow&\downarrow\\
\text{fermionic }(5,4):&1&(-1)^F&(-1)^FR&R
\end{array}. \]

\subsection{Fermionic $(9,4)$ minimal model}

Next, let us consider the fermionic $(9,4)$ minimal model with $c=-\frac{19}6$ and $c_\text{eff}=\frac56\approx0.83$. We will show that the renormalization group flow to the fermionic $(5,4)$ minimal model is inconsistent. Indeed, in \cite{Dorey:2000zb} (see also \cite{Martins:1992ht,Martins:1992yk,Ravanini:1994pt}
), it was argued that the renormalization group flow of the bosonic $(9,4)$  minimal model induced by the least relevant operator $\phi_{(1,5)}\bar{\phi}_{(1,5)}$ which is singlet  under the (hidden) $\mathbb{Z}_2$ symmetry leads to the bosonic $(7,4)$ minimal model, which has only one relevant singlet operator. Then the subsequent deformations by the singlet relevant operator will make the theory gapped. We expect that the same renormalization group flow of the fermionic cousins should exist. Note that this scenario suggests that there is a non-trivial topological field theory that ensures the anomalous chiral $\mathbb{Z}_2$ symmetry that is preserved along the renormalization group flow.

\underline{Primary operator and topological defect lines}

We first  list  primary operators in the Neveu-Schwartz sector and corresponding (Verlinde-like) topological defect lines:
\begin{align*}
    \phi_{(1,1)}\bar\phi_{(1,1)}=\phi_{(3,8)}\bar\phi_{(3,8)}&\leftrightarrow id_{0,0}\leftrightarrow1,\\
    \phi_{(3,1)}\bar\phi_{(3,1)}=\phi_{(1,8)}\bar\phi_{(1,8)}&\leftrightarrow|\ep_{\frac72}|^2\leftrightarrow(-1)^F,\\
    \phi_{(1,2)}\bar\phi_{(1,2)}=\phi_{(3,7)}\bar\phi_{(3,7)}&\leftrightarrow|\ep_{-\frac16}|^2\leftrightarrow\mcal L_{-\frac16,-\frac16},\\
    \phi_{(3,2)}\bar\phi_{(3,2)}=\phi_{(1,7)}\bar\phi_{(1,7)}&\leftrightarrow|\ep_{\frac73}|^2\leftrightarrow\mcal L_{\frac73,\frac73},\\
    \phi_{(1,3)}\bar\phi_{(1,3)}=\phi_{(3,6)}\bar\phi_{(3,6)}&\leftrightarrow|\ep_{-\frac19}|^2\leftrightarrow\mcal L_{-\frac19,-\frac19},\\
    \phi_{(3,3)}\bar\phi_{(3,3)}=\phi_{(1,6)}\bar\phi_{(1,6)}&\leftrightarrow|\ep_{\frac{25}{18}}|^2\leftrightarrow\mcal L_{\frac{25}{18},\frac{25}{18}},\\
    \phi_{(1,4)}\bar\phi_{(1,4)}=\phi_{(3,5)}\bar\phi_{(3,5)}&\leftrightarrow|\ep_{\frac16}|^2\leftrightarrow\mcal L_{\frac16,\frac16},\\
    \phi_{(3,4)}\bar\phi_{(3,4)}=\phi_{(1,5)}\bar\phi_{(1,5)}&\leftrightarrow|\ep_{\frac23}|^2\leftrightarrow\mcal L_{\frac23,\frac23},\\
    \phi_{(1,1)}\bar\phi_{(3,1)}&\leftrightarrow\bar\ep_{\frac72},\\
    \phi_{(3,1)}\bar\phi_{(1,1)}&\leftrightarrow\ep_{\frac72},\\
    \phi_{(1,2)}\bar\phi_{(3,2)}&\leftrightarrow\ep_{-\frac16}\bar\ep_{\frac73},\\
    \phi_{(3,2)}\bar\phi_{(1,2)}&\leftrightarrow\ep_{\frac73}\bar\ep_{-\frac16},\\
    \phi_{(1,3)}\bar\phi_{(3,3)}&\leftrightarrow\ep_{-\frac19}\bar\ep_{\frac{25}{18}},\\
    \phi_{(3,3)}\bar\phi_{(1,3)}&\leftrightarrow\ep_{\frac{25}{18}}\bar\ep_{-\frac19},\\
    \phi_{(1,4)}\bar\phi_{(3,4)}&\leftrightarrow\ep_{\frac16}\bar\ep_{\frac23},\\
    \phi_{(3,4)}\bar\phi_{(1,4)}&\leftrightarrow\ep_{\frac23}\bar\ep_{\frac16}.
\end{align*}
The theory has a fermionic chiral current of spin $\frac{7}{2}$.

\underline{Action of topological defect lines on primaries}\\
Solving the (modified) Cardy conditions, we find 16 topological defect lines (eight m-type generated by $\{(-1)^F,\mcal L_{-\frac16,-\frac16},\mcal L_{-\frac19,-\frac19},\mcal L_{\frac16,\frac16}\}$, and eight q-type given by the product of eight m-type and $\mcal L$) with actions
\begin{table}[H]
\hspace{-55pt}
\scalebox{0.63}{\begin{tabular}{c||c|c|c|c|c|c|c|c||c|c|c|c|c|c|c|c}
NS-NS&$id$&$|\ep_{\frac72}|^2$&$|\ep_{-\frac16}|^2$&$|\ep_{\frac73}|^2$&$|\ep_{-\frac19}|^2$&$|\ep_{\frac{25}{18}}|^2$&$|\ep_{\frac16}|^2$&$|\ep_{\frac23}|^2=|\phi_{(1,5)}|^2$&$\bar\ep_{\frac72}$&$\ep_{\frac72}$&$\ep_{-\frac16}\bar\ep_{\frac73}$&$\ep_{\frac73}\bar\ep_{-\frac16}$&$\ep_{-\frac19}\bar\ep_{\frac{25}{18}}$&$\ep_{\frac{25}{18}}\bar\ep_{-\frac19}$&$\ep_{\frac16}\bar\ep_{\frac23}$&$\ep_{\frac23}\bar\ep_{\frac16}$\\\hline
$(-1)^F$&$1$&$1$&$1$&$1$&$1$&$1$&$1$&$1$&$-1$&$-1$&$-1$&$-1$&$-1$&$-1$&$-1$&$-1$\\\hline
$\mcal L$&$-1$&$1$&$1$&$-1$&$-1$&$1$&$1$&$-1$&$1$&$-1$&$-1$&$1$&$1$&$-1$&$-1$&$1$\\\hline
$\mcal L_{-\frac16,-\frac16}$&$-2\sin\frac\pi{18}$&$-2\sin\frac\pi{18}$&$2\cos\frac\pi9$&$2\cos\frac\pi9$&$1$&$1$&$-2\cos\frac{2\pi}9$&$-2\cos\frac{2\pi}9$&$-2\sin\frac\pi{18}$&$-2\sin\frac\pi{18}$&$2\cos\frac\pi9$&$2\cos\frac\pi9$&$1$&$1$&$-2\cos\frac{2\pi}9$&$-2\cos\frac{2\pi}9$\\\hline
$\mcal L_{-\frac19,-\frac19}$&$-\frac{\sqrt3}{2\cos\frac\pi{18}}$&$-\frac{\sqrt3}{2\cos\frac\pi{18}}$&$\frac{\sqrt3}{2\sin\frac\pi9}$&$\frac{\sqrt3}{2\sin\frac\pi9}$&$0$&$0$&$\frac{\sqrt3}{2\sin\frac{2\pi}9}$&$\frac{\sqrt3}{2\sin\frac{2\pi}9}$&$-\frac{\sqrt3}{2\cos\frac\pi{18}}$&$-\frac{\sqrt3}{2\cos\frac\pi{18}}$&$\frac{\sqrt3}{2\sin\frac\pi9}$&$\frac{\sqrt3}{2\sin\frac\pi9}$&$0$&$0$&$\frac{\sqrt3}{2\sin\frac{2\pi}9}$&$\frac{\sqrt3}{2\sin\frac{2\pi}9}$\\\hline
$\mcal L_{\frac16,\frac16}$&$1-2\sin\frac\pi{18}$&$1-2\sin\frac\pi{18}$&$\frac1{2\sin\frac\pi{18}}$&$\frac1{2\sin\frac\pi{18}}$&$-1$&$-1$&$-\frac1{2\cos\frac\pi9}$&$-\frac1{2\cos\frac\pi9}$&$1-2\sin\frac\pi{18}$&$1-2\sin\frac\pi{18}$&$\frac1{2\sin\frac\pi{18}}$&$\frac1{2\sin\frac\pi{18}}$&$-1$&$-1$&$-\frac1{2\cos\frac\pi9}$&$-\frac1{2\cos\frac\pi9}$
\end{tabular}.}
\end{table}
{The $\mathcal{L}$ line is invertible and it generates the chiral $\mathbb{Z}_2$ symmetry. From the action, we learn that the relevant deformation $|\phi_{(1,5)}|^2=|\ep_{\frac23}|^2$ (as well as $|\epsilon_{-\frac{1}{9}}|^2$) preserves four topological defect lines $\{1,(-1)^F,\mcal L,(-1)^F\mcal L\}$.
We, however, emphasize that the action of $\mathcal{L}$ on the vacuum state is $-1$ rather than $+1$.  Nevertheless it is possible to derive the expected Ward identities on correlation functions. See Appendix \ref{Z2} for discussions. From this fact alone, we may argue that the $\mathcal{L}$ line cannot flow to the $R$ line of the fermionic $(5,4)$ minimal mode because action on the vacuum state (i.e. quantum dimension) is different.}

\underline{Spin contents}

By performing the modular S-transformation, we can read the spin contents of the surviving topological defect lines:
\begin{align*}
    \mcal H_{(-1)^F}:&\quad s\in\{0\},\\
    \mcal H_{\mcal L}:&\quad s\in\{-\frac{19}{16},-\frac{11}{16},-\frac3{16},\frac5{16},\frac{13}{16},\frac{21}{16},\frac{29}{16},\frac{37}{16}\}\\
    &~~~~=\{-\frac3{16},\frac5{16}\}\text{ mod }1,\\
    \mcal H_{(-1)^F\mcal L}:&\quad s\in\{-\frac{37}{16},-\frac{29}{16},-\frac{21}{16},-\frac{13}{16},-\frac5{16},\frac3{16},\frac{11}{16},\frac{19}{16}\}\\
    &~~~~=\{-\frac5{16},\frac3{16}\}\text{ mod }1.
\end{align*}
The surviving topological defect lines $\mcal L$ and $(-1)^F\mathcal{L}$ generating chiral symmetries have spin contents different from those of fermionic minimal models with $m=3,4$. Thus, the infrared theory cannot be fermionic minimal models with $m=3,4$. {We can check that they agree with that of the fermionic $(7,4)$ minimal model, and we expect it is the end of the renormalization group flow.}

\subsection{Fermionic $(E_6,A_6)$ minimal model}
{Let us now turn to exceptional cases. We will show that the fermionic $(E_6,A_6)$ minimal model with $c=-\frac{11}{14}$ and $c_\text{eff}=\frac{13}{14}\approx0.928$ (i.e. fermionic version of E-type modular invariant $(12,7)$ minimal model) cannot flow to the fermionic $(5,4)$ minimal model by a singlet deformation. We do not know the fate of the fermionic $(E_6,A_6)$ model under the relevant deformation. We conjecture that it will eventually become a non-trivial topological field theory. Note that it could flow to the fermionic $(7,4)$ minimal model (or possibly the fermionic $(9,4)$ minimal model although the number of relevant deformations does not decrease then) from the spin contents that we will  discuss below.} 

\underline{Primary operator and topological defect lines}

We list primary operators in the Neveu-Schwartz sector and (Verlinde-like) topological defect lines in terms of the extended character:
\begin{align*}
    |C_{1,1}|^2\leftrightarrow id_0&\leftrightarrow1,\\
    |C_{2,1}|^2\leftrightarrow|\ep_{\frac{11}{14}}|^2&\leftrightarrow\mcal L_{\frac{11}{14}},\\
    |C_{3,1}|^2\leftrightarrow|\ep_{\frac37}|^2&\leftrightarrow\mcal L_{\frac37},\\
    |C_{1,5}|^2\leftrightarrow|\ep_{\frac32}|^2&\leftrightarrow (-1)^F,\\
    |C_{2,5}|^2\leftrightarrow|\ep_{\frac27}|^2&\leftrightarrow\mcal L_{\frac27}=(-1)^F\mcal L_{\frac{11}{14}},\\
    |C_{3,5}|^2\leftrightarrow|\ep_{-\frac1{14}}|^2&\leftrightarrow\mcal L_{-\frac1{14}}=(-1)^F\mcal L_{\frac{17}7},\\
    C_{1,1}\bar C_{1,5}\leftrightarrow\bar\ep_{\frac32}&,\\
    C_{1,5}\bar C_{1,1}\leftrightarrow\ep_{\frac32}&,\\
    C_{2,1}\bar C_{2,5}\leftrightarrow\ep_{\frac{11}{14}}\bar\ep_{\frac27}&,\\
    C_{2,5}\bar C_{2,1}\leftrightarrow\ep_{\frac27}\bar\ep_{\frac{11}{14}},\\
    C_{3,1}\bar C_{3,5}\leftrightarrow\ep_{\frac37}\bar\ep_{-\frac1{14}},\\
    C_{3,5}\bar C_{3,1}\leftrightarrow\ep_{-\frac1{14}}\ep_{\frac37}.
\end{align*}
{We note that the fermionic $(E_6,A_6)$ minimal model has a conserved spin $\frac{3}{2}$ current, indicating that it is supersymmetric rather than non-supersymmetric. It has the same central charge as the  $\mathcal{N}=1$ supersymmetric $(7,3)$ minimal model.}

\underline{Action of topological defect lines on primaries}\\
Solving the (modified) Cardy conditions, we find six m-type topological defect lines (generated by $\{\mcal L_{\frac{11}{14}},\mcal L_{\frac37},(-1)^F\}$) and six q-type topological defect lines (fusion of $\mcal L$ with the six m-type lines) with action
\begin{table}[H]
\hspace{-30pt}
\scalebox{0.8}{\begin{tabular}{c||c|c|c|c|c|c||c|c|c|c|c|c}
NS-NS&$|\ep_0|^2$&$|\ep_{\frac{11}{14}}|^2$&$|\ep_{\frac37}|^2$&$|\ep_{\frac32}|^2$&$|\ep_{\frac27}|^2$&$|\ep_{-\frac1{14}}|^2$&$\bar\ep_{\frac32}$&$\ep_{\frac32}$&$\ep_{\frac{11}{14}}\bar\ep_{\frac27}$&$\ep_{\frac27}\bar\ep_{\frac{11}{14}}$&$\ep_{\frac37}\bar\ep_{-\frac1{14}}$&$\ep_{-\frac1{14}}\bar\ep_{\frac37}$\\\hline
$(-1)^F$&$1$&$1$&$1$&$1$&$1$&$1$&$-1$&$-1$&$-1$&$-1$&$-1$&$-1$\\
$\mcal L$&$-1$&$1$&$-1$&$1$&$-1$&$1$&$-1$&$1$&$1$&$-1$&$-1$&$1$\\
$\mcal L_{\frac{11}{14}}$&$-2\sin\frac{3\pi}{14}$&$2\sin\frac\pi{14}$&$\frac{\cos\frac{3\pi}{14}}{\sin\frac\pi7}$&$-2\sin\frac{3\pi}{14}$&$2\sin\frac\pi{14}$&$\frac{\cos\frac{3\pi}{14}}{\sin\frac\pi7}$&$-2\sin\frac{3\pi}{14}$&$2\sin\frac\pi{14}$&$\frac{\cos\frac{3\pi}{14}}{\sin\frac\pi7}$&$-2\sin\frac{3\pi}{14}$&$2\sin\frac\pi{14}$&$\frac{\cos\frac{3\pi}{14}}{\sin\frac\pi7}$\\
$\mcal L_{\frac37}$&$\frac{\sin\frac\pi7}{\cos\frac{3\pi}{14}}$&$-\frac1{2\sin\frac{3\pi}{14}}$&$\frac1{2\sin\frac\pi{14}}$&$\frac{\sin\frac\pi7}{\cos\frac{3\pi}{14}}$&$-\frac1{2\sin\frac{3\pi}{14}}$&$\frac1{2\sin\frac\pi{14}}$&$\frac{\sin\frac\pi7}{\cos\frac{3\pi}{14}}$&$-\frac1{2\sin\frac{3\pi}{14}}$&$\frac1{2\sin\frac\pi{14}}$&$\frac{\sin\frac\pi7}{\cos\frac{3\pi}{14}}$&$-\frac1{2\sin\frac{3\pi}{14}}$&$\frac1{2\sin\frac\pi{14}}$
\end{tabular}.}
\end{table}
The $\mathcal{L}$ line is invertible and it generates the chiral $\mathbb{Z}_2$ symmetry. The relevant deformations $|\ep_{\frac37}|^2,|\ep_{\frac27}|^2$ preserve four topological defect  lines $\{1, (-1)^F, \mathcal{L}, (-1)^F \mathcal{L} \}$. The action of $\mathcal{L}$ on the vacuum is $-1$ rather than $+1$, which shares the same property as that in the fermionic $(9,4)$ minimal model.  We may argue that the $\mathcal{L}$ line cannot flow to the $R$ line of the fermionic $(5,4)$ minimal mode because action on the vacuum state (i.e. quantum dimension) is different.

\underline{Spin content}

By performing the modular S-transformation, we can read the spin contents of the surviving topological defect lines:
\begin{align*}
    \mcal H_{(-1)^F}:&\quad s\in\{0\},\\
    \mcal H_{\mcal L}:&\quad s\in\{-\frac{13}{16},-\frac5{16},\frac3{16},\frac{11}{16}\}=\{-\frac5{16},\frac3{16}\}\text{ mod }1,\\
    \mcal H_{\mcal L_{\frac{11}{14}}}:&\quad s\in\{0,\pm\frac17,\pm\frac27,\pm\frac5{14},\pm\frac12,\pm\frac57,\pm\frac{11}{14},\pm\frac67,\pm\frac{17}{14}\}=\{0,\pm\frac17,\pm\frac3{14},\pm\frac27,\pm\frac5{14},\pm\frac12\}\text{ mod }1,\\
    \mcal H_{\mcal L_{\frac37}}:&\quad s\in\{0,\pm\frac1{14},\pm\frac17,\pm\frac5{14},\pm\frac37,\pm\frac12,\pm\frac67,\pm\frac{15}{14},\pm\frac{11}7\}=\{0,\pm\frac1{14},\pm\frac17,\pm\frac5{14},\frac37,\pm\frac12\}\text{ mod }1.
\end{align*}
The surviving topological lines $\mathcal{L}$ and $(-1)^F\mathcal{L}$ have the spin contents that are different from those of the fermionic $(5,4)$ minimal model, so we can rule out the possibility that it flows to the fermionic $(5,4)$ minimal model by the singlet deformation.

\subsection{Fermionic $(E_6,A_{10})$ minimal mode}\label{E6A10}

Finally, let us consider the fermionic $(E_6,A_{10})$ minimal model with $c = c_{\text{eff}}=\frac{21}{22}$ (i.e. fermionic version of E-type modular invariant $(12,11)$ minimal model) .  This is the only candidate with the unitary spectrum so it will be of interest if it is a final candidate of the non-supersymmetric  Gross-Neveu-Yukawa fixed point with a two-component Majorana fermion. We will see that the putative flow passes the constraint from the spin content of the preserved topological defect lines, and we claim it is another good candidate for the non-supersymmetric Gross-Neveu-Yukawa fixed point. See, however,  potential issues discussed in section 5.5.

\underline{Primary operator and topological line defects}

We list primary operators in the Neveu-Schwartz sector and (Verlinde-like) topological defect lines in terms of the extended character:
\begin{align*}
    C_{1,1}\bar C_{1,1}\leftrightarrow id_{0,0}&\leftrightarrow1,\\
    C_{3,1}\bar C_{3,1}\leftrightarrow\ep_{\frac{13}{11}}\bar\ep_{\frac{13}{11}}&\leftrightarrow\mcal L_{\frac{13}{11},\frac{13}{11}},\\
    C_{5,1}\bar C_{5,1}\leftrightarrow\ep_{\frac6{11}}\bar\ep_{\frac6{11}}&\leftrightarrow\mcal L_{\frac6{11},\frac6{11}},\\
  { C_{4,5}\bar C_{4,5}} \leftrightarrow\ep_{\frac1{11}}\bar\ep_{\frac1{11}}&\leftrightarrow\mcal L_{\frac1{11},\frac1{11}},\\
   {C_{2,5}\bar C_{2,5}} \leftrightarrow\ep_{\frac{20}{11}}\bar\ep_{\frac{20}{11}}&\leftrightarrow\mcal L_{\frac{20}{11},\frac{20}{11}},\\
    C_{1,5}\bar C_{1,5}\leftrightarrow\ep_{\frac72}\bar\ep_{\frac72}&\leftrightarrow(-1)^F,\\
    C_{3,5}\bar C_{3,5}\leftrightarrow\ep_{\frac{15}{22}}\bar\ep_{\frac{15}{22}}&\leftrightarrow\mcal L_{\frac{15}{22},\frac{15}{22}}=(-1)^F\mcal L_{\frac{13}{11},\frac{13}{11}},\\
    C_{5,5}\bar C_{5,5}\leftrightarrow\ep_{\frac1{22}}\bar\ep_{\frac1{22}}&\leftrightarrow\mcal L_{\frac1{22},\frac1{22}}=(-1)^F\mcal L_{\frac6{11},\frac6{11}},\\
    {C_{4,1}\bar C_{4,1}}\leftrightarrow\ep_{\frac{35}{22}}\bar\ep_{\frac{35}{22}}&\leftrightarrow\mcal L_{\frac{35}{22},\frac{35}{22}}=(-1)^F\mcal L_{\frac1{11},\frac1{11}},\\
 {C_{2,1}\bar C_{2,1}} \leftrightarrow\ep_{\frac7{22}}\bar\ep_{\frac7{22}}&\leftrightarrow\mcal L_{\frac7{22},\frac7{22}}=(-1)^F\mcal L_{\frac{20}{11},\frac{20}{11}},\\
    C_{1,1}\bar C_{1,5}\leftrightarrow\bar\ep_{\frac72}&,\\
    C_{1,5}\bar C_{1,1}\leftrightarrow\ep_{\frac72}&,\\
    C_{3,1}\bar C_{3,5}\leftrightarrow\ep_{\frac{13}{11}}\bar\ep_{\frac{15}{22}}&,\\
    C_{3,5}\bar C_{3,1}\leftrightarrow\ep_{\frac{15}{22}}\bar\ep_{\frac{13}{11}}&,\\
    C_{5,1}\bar C_{5,5}\leftrightarrow\ep_{\frac6{11}}\bar\ep_{\frac1{22}}&,\\
    C_{5,5}\bar C_{5,1}\leftrightarrow\ep_{\frac1{22}}\bar\ep_{\frac6{11}}&,\\
  {C_{4,5}\bar C_{4,1}}\leftrightarrow\ep_{\frac1{11}}\bar\ep_{\frac{35}{22}}&,\\
   {C_{4,1}\bar C_{4,5}}\leftrightarrow\ep_{\frac{35}{22}}\bar\ep_{\frac1{11}}&,\\
 {C_{2,5}\bar C_{2,1}}\leftrightarrow\ep_{\frac{20}{11}}\bar\ep_{\frac7{22}}&,\\
{C_{2,1}\bar C_{2,5}}\leftrightarrow\ep_{\frac7{22}}\bar\ep_{\frac{20}{11}}&.
\end{align*}
Note that it has a chiral fermionic current with spin $\frac{7}{2}$.

\underline{Actions of topological line defects on primaries}\\
Solving the (modified) Cardy conditions, we find 20 topological line defects (10 m-type and 10 q-type) with actions
\begin{table}[H]
\hspace{-60pt}
\scalebox{0.5}{\begin{tabular}{c||c|c|c|c|c|c|c|c|c|c||c|c|c|c|c|c|c|c|c|c}
	NS-NS&$id$&$|\ep_{\frac{13}{11}}|^2$&$|\ep_{\frac6{11}}|^2$&$|\ep_{\frac1{11}}|^2$&$|\ep_{\frac{20}{11}}|^2$&$|\ep_{\frac72}|^2$&$|\ep_{\frac{15}{22}}|^2$&$|\ep_{\frac1{22}}|^2$&$|\ep_{\frac{35}{22}}|^2$&$|\ep_{\frac7{22}}|^2$&$\bar\ep_{\frac72}$&$\ep_{\frac72}$&$\ep_{\frac{13}{11}}\bar\ep_{\frac{15}{22}}$&$\ep_{\frac{15}{22}}\bar\ep_{\frac{13}{11}}$&$\ep_{\frac6{11}}\bar\ep_{\frac1{22}}$&$\ep_{\frac1{22}}\bar\ep_{\frac6{11}}$&$\ep_{\frac1{11}}\bar\ep_{\frac{35}{22}}$&$\ep_{\frac{35}{22}}\bar\ep_{\frac1{11}}$&$\ep_{\frac{20}{11}}\bar\ep_{\frac7{22}}$&$\ep_{\frac7{22}}\bar\ep_{\frac{20}{11}}$\\\hline
	$(-1)^F$&$1$&$1$&$1$&$1$&$1$&$1$&$1$&$1$&$1$&$1$&$-1$&$-1$&$-1$&$-1$&$-1$&$-1$&$-1$&$-1$&$-1$&$-1$\\
	$\mcal L$&$1$&$1$&$1$&$1$&$1$&$-1$&$-1$&$-1$&$-1$&$-1$&$1$&$-1$&$1$&$-1$&$1$&$-1$&$1$&$-1$&$1$&$-1$\\
	$\mcal L_{\frac{13}{11},\frac{13}{11}}$&$\frac{\cos\frac{5\pi}{22}}{\sin\frac\pi{11}}$&$\frac{\sin\frac{2\pi}{11}}{\cos\frac{5\pi}{22}}$&$1-2\cos\frac\pi{11}$&$-\frac{\sin\frac\pi{11}}{\cos\frac{3\pi}{22}}$&$\frac{\cos\frac\pi{22}}{\sin\frac{2\pi}{11}}$&$\frac{\cos\frac{5\pi}{22}}{\sin\frac\pi{11}}$&$\frac{\sin\frac{2\pi}{11}}{\cos\frac{5\pi}{22}}$&$1-2\cos\frac\pi{11}$&$-\frac{\sin\frac\pi{11}}{\cos\frac{3\pi}{22}}$&$\frac{\cos\frac\pi{22}}{\sin\frac{2\pi}{11}}$&$\frac{\cos\frac{5\pi}{22}}{\sin\frac\pi{11}}$&$\frac{\cos\frac{5\pi}{22}}{\sin\frac\pi{11}}$&$\frac{\sin\frac{2\pi}{11}}{\cos\frac{5\pi}{22}}$&$\frac{\sin\frac{2\pi}{11}}{\cos\frac{5\pi}{22}}$&$1-2\cos\frac\pi{11}$&$1-2\cos\frac\pi{11}$&$-\frac{\sin\frac\pi{11}}{\cos\frac{3\pi}{22}}$&$-\frac{\sin\frac\pi{11}}{\cos\frac{3\pi}{22}}$&$\frac{\cos\frac\pi{22}}{\sin\frac{2\pi}{11}}$&$\frac{\cos\frac\pi{22}}{\sin\frac{2\pi}{11}}$\\
	$\mcal L_{\frac6{11},\frac6{11}}$&$\frac1{2\sin\frac\pi{22}}$&$-\frac1{2\sin\frac{3\pi}{22}}$&$\frac1{2\sin\frac{5\pi}{22}}$&$-\frac{\sin\frac{2\pi}{11}}{\cos\frac{3\pi}{22}}$&$\frac1{2\cos\frac\pi{11}}$&$\frac1{2\sin\frac\pi{22}}$&$-\frac1{2\sin\frac{3\pi}{22}}$&$\frac1{2\sin\frac{5\pi}{22}}$&$-\frac{\sin\frac{2\pi}{11}}{\cos\frac{3\pi}{22}}$&$\frac1{2\cos\frac\pi{11}}$&$\frac1{2\sin\frac\pi{22}}$&$\frac1{2\sin\frac\pi{22}}$&$-\frac1{2\sin\frac{3\pi}{22}}$&$-\frac1{2\sin\frac{3\pi}{22}}$&$\frac1{2\sin\frac{5\pi}{22}}$&$\frac1{2\sin\frac{5\pi}{22}}$&$-\frac{\sin\frac{2\pi}{11}}{\cos\frac{3\pi}{22}}$&$-\frac{\sin\frac{2\pi}{11}}{\cos\frac{3\pi}{22}}$&$\frac1{2\cos\frac\pi{11}}$&$\frac1{2\cos\frac\pi{11}}$\\
	$\mcal L_{\frac1{11},\frac1{11}}$&$\frac{\cos\frac{3\pi}{22}}{\sin\frac\pi{11}}$&$-\frac{\sin\frac\pi{11}}{\cos\frac{5\pi}{22}}$&$-\frac{\sin\frac{2\pi}{11}}{\cos\frac\pi{22}}$&$-\frac1{1-2\cos\frac\pi{11}}$&$-\frac{\cos\frac{5\pi}{22}}{\sin\frac{2\pi}{11}}$&$\frac{\cos\frac{3\pi}{22}}{\sin\frac\pi{11}}$&$-\frac{\sin\frac\pi{11}}{\cos\frac{5\pi}{22}}$&$-\frac{\sin\frac{2\pi}{11}}{\cos\frac\pi{22}}$&$-\frac1{1-2\cos\frac\pi{11}}$&$-\frac{\cos\frac{5\pi}{22}}{\sin\frac{2\pi}{11}}$&$\frac{\cos\frac{3\pi}{22}}{\sin\frac\pi{11}}$&$\frac{\cos\frac{3\pi}{22}}{\sin\frac\pi{11}}$&$-\frac{\sin\frac\pi{11}}{\cos\frac{5\pi}{22}}$&$-\frac{\sin\frac\pi{11}}{\cos\frac{5\pi}{22}}$&$-\frac{\sin\frac{2\pi}{11}}{\cos\frac\pi{22}}$&$-\frac{\sin\frac{2\pi}{11}}{\cos\frac\pi{22}}$&$-\frac1{1-2\cos\frac\pi{11}}$&$-\frac1{1-2\cos\frac\pi{11}}$&$-\frac{\cos\frac{5\pi}{22}}{\sin\frac{2\pi}{11}}$&$-\frac{\cos\frac{5\pi}{22}}{\sin\frac{2\pi}{11}}$\\
	$\mcal L_{\frac{20}{11},\frac{20}{11}}$&$2\cos\frac\pi{11}$&$2\sin\frac{5\pi}{22}$&$2\sin\frac\pi{22}$&$-2\sin\frac{3\pi}{22}$&$-\frac{\cos\frac{3\pi}{22}}{\sin\frac{2\pi}{11}}$&$2\cos\frac\pi{11}$&$2\sin\frac{5\pi}{22}$&$2\sin\frac\pi{22}$&$-2\sin\frac{3\pi}{22}$&$-\frac{\cos\frac{3\pi}{22}}{\sin\frac{2\pi}{11}}$&$2\cos\frac\pi{11}$&$2\cos\frac\pi{11}$&$2\sin\frac{5\pi}{22}$&$2\sin\frac{5\pi}{22}$&$2\sin\frac\pi{22}$&$2\sin\frac\pi{22}$&$-2\sin\frac{3\pi}{22}$&$-2\sin\frac{3\pi}{22}$&$-\frac{\cos\frac{3\pi}{22}}{\sin\frac{2\pi}{11}}$&$-\frac{\cos\frac{3\pi}{22}}{\sin\frac{2\pi}{11}}$
\end{tabular}.}
\end{table}
The $\mathcal{L}$ line is invertible and it generates the chiral $\mathbb{Z}_2$ symmetry. From the action, we learn that the relevant operators $|\ep_{\frac6{11}}|^2,|\ep_{\frac1{11}}|^2$ preserve four topological defect lines $\{1, (-1)^F, \mathcal{L}, (-1)^F \mathcal{L} \}$. 

\underline{Spin contents}

Performing the modular $S$-transformation, we can read the spin contents of the preserved topological defect lines:
\begin{align*}
    \mcal H_{(-1)^F}:&\quad s\in\{0\},\\
    \mcal H_{\mcal L_{\frac{13}{11},\frac{13}{11}}}:&\quad s\in\{0,\pm\frac1{22},\pm\frac3{22},\pm\frac5{22},\pm\frac5{11},\pm\frac12,\pm\frac7{11},\pm\frac{15}{22},\\
    &\hspace{30pt}\pm\frac{23}{22},\pm\frac{25}{22},\pm\frac{13}{11},\pm\frac{14}{11},\pm\frac32,\pm\frac{17}{11},\pm\frac{19}{11},\pm\frac{51}{22},\pm\frac{31}{11}\}\\
    &\hspace{50pt}=\{0,\pm\frac1{22},\pm\frac3{22},\pm\frac2{11},\pm\frac5{22},\pm\frac3{11},\pm\frac7{22},\pm\frac4{11},\pm\frac5{11},\pm\frac12\}\text{ mod }1,\\
    \mcal H_{\mcal L_{\frac6{11},\frac6{11}}}:&\quad s\in\{0,\pm\frac1{22},\pm\frac3{22},\pm\frac5{22}, \pm\frac3{11},\pm\frac9{22},\pm\frac5{11},\pm\frac12,\pm\frac6{11},\pm\frac{13}{22},\pm\frac7{11},\pm\frac{10}{11},\\
    &\hspace{30pt}\pm\frac{23}{22},\pm\frac{12}{11},\pm\frac{25}{22},\pm\frac{14}{11},\pm\frac32,\pm\frac{17}{11},\pm\frac{19}{11},\pm\frac{39}{22},\pm\frac{65}{22},\pm\frac{38}{11}\}\\
    &\hspace{50pt}=\{0,\pm\frac1{22},\pm\frac1{11},\pm\frac3{22},\pm\frac5{22},\pm\frac3{11},\pm\frac4{11},\pm\frac9{22},\pm\frac5{11},\pm\frac12\}\text{ mod }1,\\
    \mcal H_{\mcal L_{\frac1{11},\frac1{11}}}:&\quad s\in\{0,\pm\frac1{22},\pm\frac1{11},\pm\frac3{22},\pm\frac5{22},\pm\frac3{11},\pm\frac4{11},\pm\frac9{22},\pm\frac5{11},\pm\frac12,\pm\frac{13}{22},\pm\frac7{11},\pm\frac{19}{22},\pm\frac{10}{11},\\
    &\hspace{30pt}\pm\frac{23}{22},\pm\frac{12}{11},\pm\frac{25}{22},\pm\frac{14}{11},\pm\frac32,\pm\frac{17}{11},\pm\frac{35}{22},\pm\frac{39}{22},\pm\frac{21}{11},\pm\frac{75}{22}\}\\
    &\hspace{50pt}=\{0,\pm\frac1{22},\pm\frac1{11},\pm\frac3{22},\pm\frac5{22},\pm\frac3{11},\pm\frac4{11},\pm\frac9{22},\pm\frac5{11},\pm\frac12\}\text{ mod }1,\\
    \mcal H_{\mcal L_{\frac{20}{11},\frac{20}{11}}}:&\quad s\in\{0,\pm\frac1{22},\pm\frac7{22},\pm\frac4{11},\pm\frac9{22},\pm\frac5{11},\pm\frac12,\pm\frac{13}{22},\pm\frac7{11},\pm\frac{19}{22},\pm\frac{10}{11},\\
    &\hspace{30pt}\pm\frac{23}{22},\pm\frac{12}{11},\pm\frac{25}{22},\pm\frac{17}{11},\pm\frac{37}{22},\pm\frac{20}{11},\pm\frac{35}{11}\}\\
    &\hspace{50pt}=\{0,\pm\frac1{22},\pm\frac1{11},\pm\frac3{22},\pm\frac2{11},\pm\frac7{22},\pm\frac4{11},\pm\frac9{22},\pm\frac5{11},\pm\frac12\}\text{ mod }1,\\
    \mcal H_{\mcal L}:&\quad s\in\{-\frac{25}{16},-\frac{17}{16},-\frac9{16},-\frac1{16},\frac7{16},\frac{31}{16}\}=\{-\frac1{16},\frac7{16}\}\text{ mod }1,\\
    \mcal H_{(-1)^F\mcal L}:&\quad s\in\{-\frac{31}{16},-\frac7{16},\frac1{16},\frac9{16},\frac{17}{16},\frac{25}{16}\}=\{-\frac7{16},\frac1{16}\}\text{ mod }1.
\end{align*}
We see that the spin contents agree with those in the fermionic $(5,4)$ minimal model.

In this scenario, we propose the following identifications of (relevant) primary operators with the Lagrangian theories under the renormalization group flow from the non-supersymmetric fixed point (i.e. fermionic $(E_6,A_{10})$ minimal model) to the supersymmetric fixed point (i.e. fermionic $(5,4)$ minimal model).

\begin{align}
\phi &\sim |\epsilon_{\frac{1}{22}}|^2\to |\epsilon_{\frac{1}{10}}|^2 \cr
\phi^2 & \sim  |\epsilon_{\frac{1}{11}}|^2\to |\epsilon_{\frac{3}{5}}|^2 , \cr
\phi^3 , \bar{\psi}\psi & \sim   |\epsilon_{\frac{7}{22}}|^2 , |L_{-1}\epsilon_{\frac{1}{22}}|^2  \to|\epsilon_{\frac{3}{2}}|^2 , |L_{-1}\epsilon_{\frac{1}{10}}|^2 
\cr
\phi^4, \phi \psi \bar{\psi} &\sim |\epsilon_{\frac{6}{11}}|^2, |\epsilon_{\frac{13}{11}}|^2 \to |L_{-2}|^2, |L_{-2} \epsilon_{\frac{3}{5}}|^2,
\cr
\psi & \sim  \epsilon_\frac{6}{11} \bar{\epsilon}_\frac{1}{22} \to \epsilon_{\frac{3}{5}} \bar{\epsilon}_\frac{1}{10} .
 \end{align}
 We see that there is one more relevant $\mathbb{Z}_2$ odd primary operator (i.e. $|\epsilon_{\frac{15}{22}}|^2$) than in the $(5,4)$ fermionic minimal model, which may be unexpected but it is not prohibited by any renormalization group argument.
 
 Matching the spin contents, we can also identify the preserved topological defect lines:
\[ \begin{array}{ccccc}
\text{fermionic } (E_{6},A_{10}):&1&(-1)^F&\mcal L&(-1)^F\mcal L\\
&\downarrow&\downarrow&\downarrow&\downarrow\\
\text{fermionic } (5,4):&1&(-1)^F&(-1)^FR&R
\end{array}. \]

Let us finally mention the ultraviolet deformation of the fermionic $(E_6,A_{10})$ minimal model. There exists a known (integrable) renormalization group flow from the $(A_{12},E_6)$ minimal model to the $(E_6,A_{10})$ minimal model approached by $|\epsilon_{\frac{13}{11}}|^2$, which preserves the chiral $\mathbb{Z}_2$ symmetry. The (fermionic) $(A_{12},E_6)$ minimal model is further studied in Appendix B, where we show that all the known consistency relations are satisfied in this renormalization group flow.

In the scenario discussed in this subsection, it seems, however, a non-trivial task to identify this renormalization group flow in terms of the Gross-Neveu-Yukawa theory with a two-component Majorana fermion with the chiral $\mathbb{Z}_2$ symmetry. We might expect that the theory with the action
\begin{align}
S = \int d^dx \left(-\frac{1}{2} \partial^\mu \phi \partial_\mu \phi + i\bar{\psi} \gamma_\mu \partial^\mu \psi - h_1 \phi^3 \bar{\psi} \psi - h_2 \phi^8 \right) \ .
\end{align}
could give a non-supersymmetric fixed point that would be identified with the fermionic $(A_{12},E_6)$ minimal model. However, according to the $c$-theorem we expect that the central charge of this model must be greater than one and it cannot be a minimal model at least perturbatively because we know that the theory with the lower powers of fields gives the Ashkin-Teller model with central charge $c=1$. See discussions around \eqref{AT}. 

{In contrast, there is no known (integrable) ultraviolet deformation of the fermionic $(11,4)$ minimal model that preserves the chiral $\mathbb{Z}_2$ symmetry. It does not mean the non-existence, but it suggests that the renormalization group flow, if any, will be outside of the fermionic minimal models.}

\subsection{Constraint from double braiding relations?}\label{dbrconst}
At this point, we have two candidates for the non-supersymmetric Gross-Neveu-Yukawa model with a two-component Majorana fermion. It is the $(11,4)$ fermionic minimal model if it is non-unitary, and it is the $(E_6,A_{10})$ fermionic minimal model if it is unitary. It would be very interesting if there were any further constraints from the renormalization group invariant associated with topological line defects to select the final candidate. In this section, we attempt to see a possible hint from the double braiding relation.

The double braiding is given by augmenting a phase to each topological defect line in the fusion ring. 
In the A-series minimal models, it is given by the twisting of the phase $e^{2\pi i h_i}$, where $h_i$ is the conformal dimensions of the primary operator associated with the Verlinde line $\mathcal{L}_i$. It is not immediately obvious if such a definition goes through fermionic minimal models by noting e.g. the chiral $\mathbb{Z}_2$ symmetry is not associated with the Verlinde line, so the following discussions apply only to the bosonic counterpart of the renormalization group flow. In our case, the constraint should be useful because the fermionization by the $\mathbb{Z}_2$ orbifold commutes with the $\mathbb{Z}_2$ invariant renormalization group flow.

{A braiding should obey the consistency condition called the hexagon axiom. It is mathematically proven that the set of solutions to the consistency condition is discrete. As a consequence, it is natural to expect that the braidings of the preserved topological defect lines show a specific behavior under the renormalization group flow.
In \cite{KKII}, it was observed that if a preserved topological defect line $L_i$ with conformal dimension $h_i$ flows to another with $\tilde{h}_i$ under the renormalization group flow, the self double braiding satisfies the condition
\begin{align}
c_{i,i} c_{i,i} = (c_{\tilde{i},\tilde{i}} c_{\tilde{i},\tilde{i}})^*.    \label{sbr}
\end{align}
In particular, if $L_i$ can fuse with itself to the identity, the double braiding relation implies $e^{4\pi i h_i} = e^{-4 \pi i \tilde{h}_i} $. This relation was demonstrated in many examples in the adjacent renormalization group flow of two minimal models.\footnote{Here, the term adjacent or consecutive is defined in terms of the (effective) central charge.} {The relation was used to explain emergent symmetries in the minimal model including bosonic \cite{KK21}, fermionic \cite{KKII}, or non-unitary ones \cite{KKfree}. Furthermore, one can offer a more conceptual understanding of the double braiding relation from the viewpoint of extended chiral algebras \cite{KKWZW}.} By extending the idea beyond the adjacent renormalization group flow, it was then conjectured that if there exist consecutive renormalization group flows that connect three renormalization group fixed points, preserving a particular topological line, the self double braiding of the preserved line (e.g. $e^{4\pi i h_i}$) must be real. }

{We, however, see that if we have a consecutive renormalization group flow --- the one from the non-supersymmetric Gross-Neveu-Yukawa fixed point to the fermionic $(5,4)$ minimal model --- ending up with the fermionic $(4,3)$ minimal model, the conjecture cannot hold because  the topological defect line that corresponds to the chiral $\mathbb{Z}_2$ symmetry (or more precisely the Kramers-Wannier duality line in the bosonic counterpart) does have a non-real double braiding.}

{Still, one may see a slight difference between the double braidings of the Kramers-Wannier duality line in the $(11,4)$ minimal model and that in the $(E_{6},A_{10})$ minimal model. Compared with the $(5,4)$ minimal model, the phase is opposite in the former and it is the same in the latter. 
Concretely, the Kramers-Wannier duality line in the bosonic $(11,4)$ minimal model has the self double braiding 
\[ c^{(11,4)}_{N,N}c^{(11,4)}_{N,N}=e^{-\frac{\pi i}4}id_0\oplus e^{\frac{3\pi i}4}id_\eta, \]
and the Kramers-Wannier duality line in the bosonic $(E_6,A_{10})$ minimal model has the self double braiding
\[ c^{(E_6,A_{10})}_{N,N}c^{(E_6,A_{10})}_{N,N}=e^{\frac{\pi i}4}id_0\oplus e^{-\frac{3\pi i}4}id_\eta, \]
while the Kramers-Wannier duality line $N$ in the bosonic $(5,4)$ minimal model has self double braiding 
\[ c^{(5,4)}_{N,N}c^{(5,4)}_{N,N}=e^{\frac{\pi i}4}id_0\oplus e^{-\frac{3\pi i}4}id_\eta. \]}

{If we assume that the {\it adjacent} renormalization group flow should give the constraint \eqref{sbr},\footnote{{We show in Appendix B.7 and B.8 that in a known (integrable) renormalization group flow between the E-series minimal models, the 
 constraint \eqref{sbr} is satisfied.}} then we may prefer the $(11,4)$ minimal model to the $(E_{6},A_{10})$ minimal model as a candidate for the non-supersymmetric Gross-Neveu-Yukawa fixed point because the putative flow from the $(E_{6},A_{10})$ minimal model to the $(5,4)$ minimal model does not satisfy \eqref{sbr}. 
 The flow that satisfies \eqref{sbr} is the one from the $(E_{6},A_{10})$ minimal model directly to the $(4,3)$ minimal model without stopping by the $(5,4)$ minimal model, and we may argue that the $(4,3)$ minimal model is the adjacent fixed point.}

\section{Discussions}

In this paper, we have argued that the non-supersymmetric fixed point of the Gross-Neveu-Yukawa model with a two-component Majorana fermion is either fermionic $(11,4)$ minimal model if non-unitary or the fermionic $(E_6,A_{10})$ minimal model if unitary.
 If we further use a constraint from the double braiding relation,
the former scenario is preferable.

The renormalization group flow from the $(11,4)$ minimal model to the $(5,4)$ minimal model has been studied in the literature \cite{Dorey:2000zb}. We have not found the study of the renormalization group flow from the $(E_6,A_{10})$ model in the literature. Once we know the structure constant of the theory, one may attempt the truncated conformal space approach. One may also use the thermodynamic Bethe ansatz to study the renormalization group flow.

Let us conclude this paper with the predictions in three dimensions. We know the perturbative computations near four dimensions \cite{Fei:2016sgs} and we have a prediction in two dimensions. We may use these data to extrapolate the critical exponents in three dimensions.

Since the fermionic $(11,4)$ minimal model seems more strongly coupled than the fermionic $(5,4)$ minimal model from the anomalous dimensions, it is not obvious if the continuation of dimensionality from $\epsilon$ expansion is reliable. We nevertheless attempt to predict the scaling dimensions of operators in three dimensions by Pad\'e approximation. 

Perturbative computation by Fei et al together with the boundary condition at $\epsilon=2$ reads
\begin{align}
\Delta_\phi &= 1 -\frac{3}{7} \epsilon -\frac{95}{6174}\epsilon^2 + O(\epsilon^3) = -\frac{5}{11} \ \ \text{at} \ \epsilon=2 \cr
\Delta_{\psi} &= \frac{3}{2} -\frac{3}{7}\epsilon - \frac{115}{37044} \epsilon^2 + O(\epsilon^3) = \frac{19}{22}  \ \ \text{at} \ \epsilon=2 \cr
\Delta_{\phi^2} & = 2 -\frac{22}{21}\epsilon + \frac{1117}{9261}  \epsilon^2  + O(\epsilon^3) = -\frac{6}{11}  \ \ \text{at} \ \epsilon=2 \ .
\end{align}
The Pad\'e $[1,2]$ extrapolation at $\epsilon=1$ gives $\Delta_{\phi} = 0.520$, $\Delta_{\psi} = 1.12 $ and $\Delta_{\phi^2} = 0.986$.  The Pad\'e $[2,1]$ extrapolation at $\epsilon=1$ gives $\Delta_{\phi} = 0.544$, $\Delta_{\psi} = 1.06 $. We have observed that $\Delta_{\phi^2}$ in the latter case has a pole at $\epsilon < 1$, so it is not reliable.

A similar analysis for the fermionic $(E_6,A_{10})$ minimal model can be done. 

\begin{align}
\Delta_\phi &= 1 -\frac{3}{7} \epsilon -\frac{95}{6174}\epsilon^2 + O(\epsilon^3) = \frac{1}{11} \ \ \text{at} \ \epsilon=2 \cr
\Delta_{\psi} &= \frac{3}{2} -\frac{3}{7}\epsilon - \frac{115}{37044} \epsilon^2 + O(\epsilon^3) = \frac{13}{22} \ \  \text{at} \ \epsilon=2 \cr
\Delta_{\phi^2} & = 2 -\frac{22}{21}\epsilon + \frac{1117}{9261}  \epsilon^2  + O(\epsilon^3) = \frac{2}{11}  \ \ \text{at} \ \epsilon=2 \ .
\end{align}
The Pad\'e $[1,2]$ extrapolation at $\epsilon=1$ gives $\Delta_{\phi} = 0.557$, $\Delta_{\psi} = 1.06 $ and $\Delta_{\phi^2} = 1.04$.  The Pad\'e $[2,1]$ extrapolation at $\epsilon=1$ gives $\Delta_{\phi} = 0.557$, $\Delta_{\psi} = 1.07 $ and  $\Delta_{\phi^2} = 1.04$. These numbers are close to the ones obtained in \cite{Fei:2016sgs}.

There have been further studies of the Gross-Neveu-Yukawa model from different approaches 
\cite{Mihaila:2017ble,Iliesiu:2017nrv,Gies:2017tod,Gracey:2017fzu,Zerf:2017zqi,Manashov:2017rrx,Erramilli:2022kgp}, but the non-supersymmetric fixed point has been rarely
studied. It would be interesting to apply these ideas to the non-supersymmetric fixed point to verify our predictions.

\section*{Acknowledgement}
NY is in part supported by JSPS KAKENHI Grant Number 21K03581.

The authors thank the Yukawa Institute for Theoretical Physics at Kyoto University, where this work was initiated during the YITP workshop ``Strings and Fields 2022".

\appendix
\setcounter{section}{0}
\renewcommand{\thesection}{\Alph{section}}
\setcounter{equation}{0}
\renewcommand{\theequation}{\Alph{section}.\arabic{equation}}

\section{Summary of Yellow book convention}
A minimal model is specified by a pair of coprime natural numbers $(p,p')$ with convention $p>p'$. (See the yellow book \cite{DiFrancesco:1997nk} p.218.) The theory has a central charge
\begin{equation}
    c=1-6\frac{(p-p')^2}{pp'}.\label{ccharge}
\end{equation}
Primary operators of the model are labeled by Kac indices
\begin{equation}
    E_{p,p'}:=\{(r,s)|1\le r\le p'-1,1\le s\le p-1\}/\sim\label{Kac}
\end{equation}
where the equivalence relation is given by
\begin{equation}
    (r,s)\sim(p'-r,p-s).\label{Kacequiv}
\end{equation}
Primary operator $\phi_{(r,s)}$ of the model has conformal dimension
\begin{align}
h_{r,s} &= \frac{(pr-p's)^2 - (p-p')^2}{4pp'}. 
\end{align}
The modular $S$-matrix of the theory is given by
\begin{align}
S_{rs,\rho\sigma} = 2\sqrt{\frac{2}{pp'}} (-1)^{1+ s \rho + r \sigma} \sin(\pi \frac{p}{p'} r \rho)\sin(\pi \frac{p'}{p} s\sigma) .
\end{align}

\section{Related minimal models}
In this appendix, we collect some properties of topological defect lines  in related minimal models. We order them according to their (effective) central charges.

\subsection{Fermionic $(4,3)$ minimal model}
This is a free Majorana fermion and the fermionic version of the Ising model with $c=\frac{1}{2}$. It is also obtained by a relevant deformation of the fermionic tricritical Ising model. It has a chiral $\mathbb{Z}_2$ symmetry.

\underline{Primary operators and Verlinde lines}

Primary states in the Neveu-Schwartz sector and the corresponding (Verlinde-like) topological defect lines are
\begin{align*}
    \phi_{(1,1)}\bar{\phi}_{(1,1)}=\phi_{(3,2)}\bar{\phi}_{(3,2)} &\leftrightarrow id_{0,0} \leftrightarrow1,\\
    \phi_{(1,2)}\bar{\phi}_{(1,2)} = \phi_{(3,1)}\bar{\phi}_{(3,1)}  &\leftrightarrow|\epsilon_{\frac{1}{2}}|^2 \leftrightarrow (-1)^F, \\
    \phi_{(1,1)}\bar{\phi}_{(1,2)}=\phi_{(3,2)}\bar{\phi}_{(3,1)} &\leftrightarrow \bar{\epsilon}_{\frac{1}{2}},\\
    \phi_{(1,2)}\bar{\phi}_{(1,1)}=\phi_{(3,1)}\bar{\phi}_{(3,2)} &\leftrightarrow \epsilon_{\frac{1}{2}}. \\
\end{align*}

\underline{Action of topological defect lines on primaries}

{We find four topological defect lines (two m-type and two q-type) with the action}

\begin{table}[H]
\begin{center}
\begin{tabular}{c|c|c||c|c}
	NS-NS&$id$&$|\ep_{\frac1{2}}|^2$&$\ep_{\frac{1}{2}}$&$\bar\ep_{\frac12}$\\\hline
	$(-1)^F$&$1$&$1$&$-1$&$-1$\\
	$(-1)^{F_L}$&$1$&$-1$&$-1$&$1$
\end{tabular}.
\end{center}
\end{table}

\underline{Spin contents}
\begin{align*}
    \mcal H_{(-1)^F}:&\quad s\in\{0\},\\
    \mcal H_{(-1)^{F_{L}}}:&\quad s\in \{-\frac7{16},\frac1{16}\},\\
    \mcal H_{(-1)^{F_{R}}}:&\quad s\in \{-\frac1{16},\frac7{16}\}.
\end{align*}

Note that the spin content is the same as the one in the fermionic $(5,4)$ minimal model, and it is consistent with the renormalization group flow from the fermionic $(5,4)$ minimal model to the fermionic $(4,3)$ minimal model by the singlet deformation.

\subsection{Bosonic $(5,4)$ minimal model}
This is a tricritical Ising model and the bosonic version of the supersymmetric Ising model with $c=\frac{7}{10}$.

\underline{Primary operators and Verlinde lines}
\begin{align*}
	\phi_{(1,1)}=\phi_{(3,4)}&\leftrightarrow id_0\leftrightarrow1,\\
	\phi_{(2,1)}=\phi_{(2,4)}&\leftrightarrow\sigma_{\frac7{16}}\leftrightarrow N,\\
	\phi_{(2,2)}=\phi_{(2,3)}&\leftrightarrow\sigma_{\frac3{80}}\leftrightarrow WN,\\
	\phi_{(3,1)}=\phi_{(1,4)}&\leftrightarrow\ep_{\frac32}\leftrightarrow\eta,\\
	\phi_{(3,2)}=\phi_{(1,3)}&\leftrightarrow\ep_{\frac35}\leftrightarrow W,\\
	\phi_{(3,3)}=\phi_{(1,2)}&\leftrightarrow\ep_{\frac1{10}}\leftrightarrow\eta W.
\end{align*}
The rank six modular tensor category is generated by three primitive lines $\{N,\eta,W\}$.

\underline{Action of topological defect lines on primaries}
\begin{table}[H]
\begin{center}
\begin{tabular}{c|c|c|c|c|c|c}
	&$id$&$\ep_{\frac1{10}}$&$\ep_{\frac35}$&$\ep_{\frac32}$&$\sigma_{\frac3{80}}$&$\sigma_{\frac7{16}}$\\\hline
	$\eta$&$1$&$1$&$1$&$1$&$-1$&$-1$\\
	$W$&$\zeta$&$-\zeta^{-1}$&$-\zeta^{-1}$&$\zeta$&$-\zeta^{-1}$&$\zeta$\\
	$N$&$\sqrt2$&$-\sqrt2$&$\sqrt2$&$-\sqrt2$&0&0
\end{tabular}.
\end{center}
\end{table}
Here, $\zeta:=\frac{1+\sqrt5}2$.

\underline{Spin content}
\begin{align*}
    \mcal H_\eta:&\quad s\in\{0,\pm\frac12,\pm\frac32\}=\{0,\pm\frac12\}\text{ mod }1,\\
    \mcal H_W:&\quad s\in\{0,\pm\frac25,\pm\frac35,\pm\frac75\}=\{0,\pm\frac25\}\text{ mod }1,\\
    \mcal H_N:&\quad s\in\{\pm\frac1{16},\pm\frac7{16},\pm\frac9{16},\pm\frac{17}{16}\}=\{\pm\frac1{16},\pm\frac7{16}\}\text{ mod }1.
\end{align*}

\subsection{Bosonic $(9,4)$ minimal model}
Here, we present some properties of topological defect lines (Verlinde lines) in the bosonic A-series $(9,4)$ minimal model, which has $c=-\frac{19}6,c_\text{eff}=\frac56\approx0.83$.

\underline{Primary operators and Verlinde lines}
\begin{align*}
    \phi_{(1,1)}=\phi_{(3,8)}&\leftrightarrow id_0\leftrightarrow1,\\
    \phi_{(2,1)}=\phi_{(2,8)}&\leftrightarrow\sigma_{\frac{19}{16}}\leftrightarrow\mcal L_{2,1}=N,\\
    \phi_{(3,1)}=\phi_{(1,8)}&\leftrightarrow\ep_{\frac72}\leftrightarrow\mcal L_{3,1}=\eta,\\
    \phi_{(1,2)}=\phi_{(3,7)}&\leftrightarrow\ep_{-\frac16}\leftrightarrow\mcal L_{1,2},\\
    \phi_{(2,2)}=\phi_{(2,7)}&\leftrightarrow\sigma_{\frac{25}{48}}\leftrightarrow\mcal L_{2,2}=N\mcal L_{1,2},\\
    \phi_{(3,2)}=\phi_{(1,7)}&\leftrightarrow\ep_{\frac73}\leftrightarrow\mcal L_{2,3}=\eta\mcal L_{1,2},\\
    \phi_{(1,3)}=\phi_{(3,6)}&\leftrightarrow\ep_{-\frac19}\leftrightarrow\mcal L_{1,3},\\
    \phi_{(2,3)}=\phi_{(2,6)}&\leftrightarrow\sigma_{\frac{11}{144}}\leftrightarrow\mcal L_{2,3}=N\mcal L_{1,3},\\
    \phi_{(3,3)}=\phi_{(1,6)}&\leftrightarrow\ep_{\frac{25}{18}}\leftrightarrow\mcal L_{3,3}=\eta\mcal L_{1,3},\\
    \phi_{(1,4)}=\phi_{(3,5)}&\leftrightarrow\ep_{\frac16}\leftrightarrow\mcal L_{1,4},\\
    \phi_{(2,4)}=\phi_{(2,5)}&\leftrightarrow\sigma_{-\frac7{48}}\leftrightarrow\mcal L_{2,4}=N\mcal L_{1,4},\\
    \phi_{(3,4)}=\phi_{(1,5)}&\leftrightarrow\ep_{\frac23}\leftrightarrow\mcal L_{1,5}=\eta\mcal L_{1,4}.
\end{align*}
Thus the rank 12 modular tensor category is generated by five primitive lines $\{N,\eta,\mcal L_{2,1},\mcal L_{3,1},\mcal L_{4,1}\}$.

\underline{Action of topological defect lines on primaries}
\begin{table}[H]
\hspace{-50pt}
\scalebox{0.9}{$\begin{tabular}{c|c|c|c|c|c|c|c|c|c|c|c|c}
	&$id$&$\sigma_{\frac{19}{16}}$&$\ep_{\frac72}$&$\ep_{-\frac16}$&$\sigma_{\frac{25}{48}}$&$\ep_{\frac73}$&$\ep_{-\frac19}$&$\sigma_{\frac{11}{144}}$&$\ep_{\frac{25}{18}}$&$\ep_{\frac16}$&$\sigma_{-\frac7{48}}$&$\ep_{\frac23}$\\\hline
	$\eta$&$1$&$-1$&$1$&$1$&$-1$&$1$&$1$&$-1$&$1$&$1$&$-1$&$1$\\\hline
	$N$&$-\sqrt2$&$0$&$\sqrt2$&$\sqrt2$&$0$&$-\sqrt2$&$-\sqrt2$&$0$&$\sqrt2$&$\sqrt2$&$0$&$-\sqrt2$\\\hline
	$\mcal L_{1,2}$&$-2\sin\frac\pi{18}$&$2\sin\frac\pi{18}$&$-2\sin\frac\pi{18}$&$2\cos\frac\pi9$&$-2\cos\frac\pi9$&$2\cos\frac\pi9$&$1$&$-1$&$1$&$-2\cos\frac{2\pi}9$&$\frac{\cos\frac\pi{18}}{\sin\frac{2\pi}9}$&$-2\cos\frac{2\pi}9$\\\hline
	$\mcal L_{1,3}$&$-\frac{\sqrt3}{2\cos\frac\pi{18}}$&$-\frac{\sqrt3}{2\cos\frac\pi{18}}$&$-\frac{\sqrt3}{2\cos\frac\pi{18}}$&$\frac{\sqrt3}{2\sin\frac\pi9}$&$\frac{\sqrt3}{2\sin\frac\pi9}$&$\frac{\sqrt3}{2\sin\frac\pi9}$&$0$&$0$&$0$&$\frac{\sqrt3}{2\sin\frac{2\pi}9}$&$\frac{\sqrt3}{2\sin\frac{2\pi}9}$&$\frac{\sqrt3}{2\sin\frac{2\pi}9}$\\\hline
	$\mcal L_{1,4}$&$1-2\sin\frac\pi{18}$&$-1+2\sin\frac\pi{18}$&$1-2\sin\frac\pi{18}$&$\frac1{2\sin\frac\pi{18}}$&$-\frac1{2\sin\frac\pi{18}}$&$\frac1{2\sin\frac\pi{18}}$&$-1$&$1$&$-1$&$-\frac1{2\cos\frac\pi9}$&$\frac1{2\cos\frac\pi9}$&$-\frac1{2\cos\frac\pi9}$
\end{tabular}$}.
\end{table}

\underline{Spin contents}
\begin{align*}
    \mcal H_\eta:&\quad s\in\{0,\pm\frac12,\pm\frac32,\pm\frac52,\pm\frac72\}=\{0,\pm\frac12\}\text{ mod }1,\\
    \mcal H_N:&\quad s\in\{\pm\frac3{16},\pm\frac5{16},\pm\frac{11}{16},\pm\frac{13}{16},\pm\frac{19}{16},\pm\frac{21}{16},\pm\frac{29}{16},\pm\frac{37}{16}\}=\{\pm\frac3{16},\pm\frac5{16}\}\text{ mod }1,\\
    \mcal H_{\mcal L_{1,2}}:&\quad s\in\{0,\pm\frac1{18},\pm\frac16,\pm\frac29,\pm\frac5{18},\pm\frac49,\pm\frac12,\pm\frac23,\pm\frac{13}{18},\pm\frac{17}{18},\pm\frac76\}\\
    &\hspace{30pt}=\{0,\pm\frac1{18},\pm\frac16,\pm\frac29,\pm\frac5{18},\pm\frac49,\pm\frac12\}\text{ mod }1,\\
    \mcal H_{\mcal L_{1,3}}:&\quad s\in\{0,\pm\frac19,\pm\frac29,\pm\frac13,\pm\frac23,\pm\frac79,\pm\frac{10}9,\pm\frac{11}9,\pm\frac53,\pm\frac{19}9\}\\
    &\hspace{30pt}=\{0,\pm\frac19,\pm\frac29,\pm\frac13\}\text{ mod }1,\\
    \mcal H_{\mcal L_{1,4}}:&\quad s\in\{0,\pm\frac1{18},\pm\frac16,\pm\frac29,\pm\frac5{18},\pm\frac49,\pm\frac12,\pm\frac23,\pm\frac{13}{18},\pm\frac56,\pm\frac{17}{18},\pm\frac43,\pm\frac32,\pm\frac{13}6,\pm\frac{17}6\}\\
    &\hspace{30pt}=\{0,\pm\frac1{18},\pm\frac16,\pm\frac29,\pm\frac5{18},\pm\frac13,\pm\frac49,\pm\frac12\}\text{ mod }1,\\
\end{align*}

Given these data, let us consider whether this ultraviolet theory can flow to the bosonic $(5,4)$ minimal model. The ultraviolet theory has the Tambara-Yamagami category, but the $N$ line has quantum dimension $-\sqrt2$. Since we assume that the renormalization group flow preserves the $N$ line, the mismatch of the quantum dimensions makes the renormalization group flow unlikely.

\subsection{Bosonic $(11,4)$ minimal model}
We consider the bosonic $(11,4)$ minimal model with $c=-\frac{125}{22},c_\text{eff}=\frac{19}{22}\approx0.86$. The aim of the following discussion is that the renormalization group flow from the bosonic $(11,4)$ minimal model to the bosonic $(5,4)$ minimal model passes all the constraints on the topological defect lines, which may be a further consistency check of the fermionic renormalization group flow studied in section 5.1.

\underline{Primary operators and Verlinde lines}
\begin{align*}
    \phi_{(1,1)}=\phi_{(3,10)}&\leftrightarrow id_0\leftrightarrow1,\\
    \phi_{(2,1)}=\phi_{(2,10)}&\leftrightarrow\sigma_{\frac{25}{16}}\leftrightarrow N,\\
    \phi_{(3,1)}=\phi_{(1,10)}&\leftrightarrow\ep_{\frac92}\leftrightarrow\eta,\\
    \phi_{(1,2)}=\phi_{(3,9)}&\leftrightarrow\ep_{-\frac5{22}}\leftrightarrow\mcal L_{1,2},\\
    \phi_{(2,2)}=\phi_{(2,9)}&\leftrightarrow\sigma_{\frac{147}{176}}\leftrightarrow\mcal L_{2,2}=N\mcal L_{1,2},\\
    \phi_{(3,2)}=\phi_{(1,9)}&\leftrightarrow\ep_{\frac{36}{11}}\leftrightarrow\mcal L_{3,2}=\eta\mcal L_{1,2},\\
    \phi_{(1,3)}=\phi_{(3,8)}&\leftrightarrow\ep_{-\frac3{11}}\leftrightarrow\mcal L_{1,3},\\
    \phi_{(2,3)}=\phi_{(2,8)}&\leftrightarrow\sigma_{\frac{51}{176}}\leftrightarrow\mcal L_{2,3}=N\mcal L_{1,3},\\
    \phi_{(3,3)}=\phi_{(1,8)}&\leftrightarrow\ep_{\frac{49}{22}}\leftrightarrow\mcal L_{3,3}=\eta\mcal L_{1,3},\\
    \phi_{(1,4)}=\phi_{(3,7)}&\leftrightarrow\ep_{-\frac3{22}}\leftrightarrow\mcal L_{1,4},\\
    \phi_{(2,4)}=\phi_{(2,7)}&\leftrightarrow\sigma_{-\frac{13}{176}}\leftrightarrow\mcal L_{2,4}=N\mcal L_{1,4},\\
    \phi_{(3,4)}=\phi_{(1,7)}&\leftrightarrow\ep_{\frac{15}{11}}\leftrightarrow\mcal L_{3,4}=\eta\mcal L_{1,4},\\
    \phi_{(1,5)}=\phi_{(3,6)}&\leftrightarrow\ep_{\frac2{11}}\leftrightarrow\mcal L_{1,5},\\
    \phi_{(2,5)}=\phi_{(2,6)}&\leftrightarrow\sigma_{-\frac{45}{176}}\leftrightarrow\mcal L_{2,5}=N\mcal L_{1,5},\\
    \phi_{(3,5)}=\phi_{(1,6)}&\leftrightarrow\ep_{\frac{15}{22}}\leftrightarrow\mcal L_{3,5}=\eta\mcal L_{1,5}.
\end{align*}
Thus the rank 15 modular tensor category is generated by six primitive lines $\{N,\eta,\mcal L_{1,2},\mcal L_{1,3},\mcal L_{1,4},\mcal L_{1,5}\}$.

\underline{Action of topological defect lines on primaries}
\begin{table}[H]
\hspace{-60pt}
\scalebox{0.69}{$\begin{tabular}{c|c|c|c|c|c|c|c|c|c|c|c|c|c|c|c}
	&$id$&$\sigma_{\frac{25}{16}}$&$\ep_{\frac92}$&$\ep_{-\frac5{22}}$&$\sigma_{\frac{147}{176}}$&$\ep_{\frac{36}{11}}$&$\ep_{-\frac3{11}}$&$\sigma_{\frac{51}{176}}$&$\ep_{\frac{49}{22}}$&$\ep_{-\frac3{22}}$&$\sigma_{-\frac{13}{176}}$&$\ep_{\frac{15}{11}}$&$\ep_{\frac2{11}}$&$\sigma_{-\frac{45}{176}}$&$\ep_{\frac{15}{22}}$\\\hline
	$\eta$&$1$&$-1$&$1$&$1$&$-1$&$1$&$1$&$-1$&$1$&$1$&$-1$&$1$&$1$&$-1$&$1$\\\hline
	$N$&$\sqrt2$&$0$&$-\sqrt2$&$-\sqrt2$&$0$&$\sqrt2$&$\sqrt2$&$0$&$-\sqrt2$&$-\sqrt2$&$0$&$\sqrt2$&$\sqrt2$&$0$&$-\sqrt2$\\\hline
	$\mcal L_{1,2}$&$-2\sin\frac{3\pi}{22}$&$2\sin\frac{3\pi}{22}$&$-2\sin\frac{3\pi}{22}$&$2\sin\frac{5\pi}{22}$&$-2\sin\frac{5\pi}{22}$&$2\sin\frac{5\pi}{22}$&$2\cos\frac\pi{11}$&$-2\cos\frac\pi{11}$&$2\cos\frac\pi{11}$&$2\sin\frac\pi{22}$&$-2\sin\frac\pi{22}$&$2\sin\frac\pi{22}$&$-\frac{\cos\frac{3\pi}{22}}{\sin\frac{2\pi}{11}}$&$\frac{\cos\frac{3\pi}{22}}{\sin\frac{2\pi}{11}}$&$-\frac{\cos\frac{3\pi}{22}}{\sin\frac{2\pi}{11}}$\\\hline
	$\mcal L_{1,3}$&$-\frac{\sin\frac\pi{11}}{\cos\frac{3\pi}{22}}$&$-\frac{\sin\frac\pi{11}}{\cos\frac{3\pi}{22}}$&$-\frac{\sin\frac\pi{11}}{\cos\frac{3\pi}{22}}$&$\frac{\sin\frac{2\pi}{11}}{\cos\frac{5\pi}{22}}$&$\frac{\sin\frac{2\pi}{11}}{\cos\frac{5\pi}{22}}$&$\frac{\sin\frac{2\pi}{11}}{\cos\frac{5\pi}{22}}$&$\frac{\cos\frac{5\pi}{22}}{\sin\frac\pi{11}}$&$\frac{\cos\frac{5\pi}{22}}{\sin\frac\pi{11}}$&$\frac{\cos\frac{5\pi}{22}}{\sin\frac\pi{11}}$&$1-2\cos\frac\pi{11}$&$1-2\cos\frac\pi{11}$&$1-2\cos\frac\pi{11}$&$\frac{\cos\frac\pi{22}}{\sin\frac{2\pi}{11}}$&$\frac{\cos\frac\pi{22}}{\sin\frac{2\pi}{11}}$&$\frac{\cos\frac\pi{22}}{\sin\frac{2\pi}{11}}$\\\hline
	$\mcal L_{1,4}$&$\frac1{-1+2\cos\frac\pi{11}}$&$-\frac{\cos\frac\pi{22}}{\cos\frac{3\pi}{22}}$&$\frac1{-1+2\cos\frac\pi{11}}$&$-\frac{\sin\frac\pi{11}}{\cos\frac{5\pi}{22}}$&$\frac{\sin\frac\pi{11}}{\cos\frac{5\pi}{22}}$&$-\frac{\sin\frac\pi{11}}{\cos\frac{5\pi}{22}}$&$\frac{\cos\frac{3\pi}{22}}{\sin\frac\pi{11}}$&$-\frac{\cos\frac{3\pi}{22}}{\sin\frac\pi{11}}$&$\frac{\cos\frac{3\pi}{22}}{\sin\frac\pi{11}}$&$-\frac{\sin\frac{2\pi}{11}}{\cos\frac\pi{22}}$&$\frac{\sin\frac{2\pi}{11}}{\cos\frac\pi{22}}$&$-\frac{\sin\frac{2\pi}{11}}{\cos\frac\pi{22}}$&$-\frac{\cos\frac{5\pi}{22}}{\sin\frac{2\pi}{11}}$&$\frac{\cos\frac{5\pi}{22}}{\sin\frac{2\pi}{11}}$&$-\frac{\cos\frac{5\pi}{22}}{\sin\frac{2\pi}{11}}$\\\hline
	$\mcal L_{1,5}$&$-\frac{\sin\frac{2\pi}{11}}{\cos\frac{3\pi}{22}}$&$-\frac{\sin\frac{2\pi}{11}}{\cos\frac{3\pi}{22}}$&$-\frac{\sin\frac{2\pi}{11}}{\cos\frac{3\pi}{22}}$&$-\frac1{2\sin\frac{3\pi}{22}}$&$-\frac1{2\sin\frac{3\pi}{22}}$&$-\frac1{2\sin\frac{3\pi}{22}}$&$\frac1{2\sin\frac\pi{22}}$&$\frac1{2\sin\frac\pi{22}}$&$\frac1{2\sin\frac\pi{22}}$&$\frac1{2\sin\frac{5\pi}{22}}$&$\frac1{2\sin\frac{5\pi}{22}}$&$\frac1{2\sin\frac{5\pi}{22}}$&$\frac1{2\cos\frac\pi{11}}$&$\frac1{2\cos\frac\pi{11}}$&$\frac1{2\cos\frac\pi{11}}$
\end{tabular}$}.
\end{table}

\underline{Quantum dimensions and braidings}

Since the fermionic counterpart of this theory is a promising candidate for the non-supersymmetric Gross-Neveu-Yukawa fixed point, we elaborate on the property of topological defect lines here. The quantum dimensions of each line can be computed as
\begin{align*}
    \hspace{-0pt}d_1=1=d_\eta,\quad d_N=\sqrt2,\quad d_{1,2}=-2\sin\frac{3\pi}{22}=d_{3,2},&\quad d_{2,2}=-2\sqrt2\sin\frac{3\pi}{22},\quad d_{1,3}=-\frac{\sin\frac\pi{11}}{\cos\frac{3\pi}{22}}=d_{3,3},\\
    d_{2,3}=-\frac{\sqrt2\sin\frac\pi{11}}{\cos\frac{3\pi}{22}},\quad d_{1,4}=\frac1{-1+2\cos\frac\pi{11}}=d_{3,4},&\quad d_{2,4}=\frac{\sqrt2}{-1+2\cos\frac\pi{11}},\\
    d_{1,5}=-\frac{\sin\frac{2\pi}{11}}{\cos\frac{3\pi}{22}}=d_{3,5},\quad& d_{2,5}=-\frac{\sqrt2\sin\frac{2\pi}{11}}{\cos\frac{3\pi}{22}}.
\end{align*}

{The global dimension $D^2:=\sum_jd_j^2$ is given by
\[ D=-\frac{\sqrt{11}}{\cos\frac{3\pi}{22}}\approx-3.65. \]}

The relevant $\phi_{(1,5)}$-deformation preserves rank three braided fusion category $\{1,\eta,N\}$, i.e., Tambara-Yamagami category. Their double braidings are given by
\[ \begin{pmatrix}id_1&id_\eta&id_N\\id_\eta&id_1&-id_N\\id_N&-id_N&e^{-\frac{\pi i}4}id_1\oplus e^{\frac{3\pi i}4}id_\eta\end{pmatrix}. \]

Since the surviving topological line defect $N$ has quantum dimension $d_N\notin\mbb N$, the infrared theory cannot be trivial. Thus, the infrared theory is either a topological quantum field theory with ground state degeneracy $>1$, or a conformal field theory. For our purpose, we assume that it is a conformal field theory.
{ We first note that the renormalization group flow from the bosonic $(11,4)$ minimal model to the bosonic $(5,4)$ minimal model proposed in \cite{Dorey:2000zb} satisfies the double braiding relation \eqref{sbr} 
 that we discussed in section \ref{dbrconst}. The renormalization group flow from the bosonic $(5,4)$ minimal model to the bosonic $(4,3)$ minimal model also satisfies the double braiding relation. However, the direct renormalization group flow from the bosonic $(11,4)$ minimal model to the bosonic $(4,3)$ minimal model does {\it not} satisfy the double braiding 
 relation, and we interpret it as not being the adjacent fixed point. To look for other possibilities of the adjacent fixed points than the bosonic $(4,3)$ minimal model, we further run the search for a consistent modular tensor category as in \cite{KK21,KKfree}.}

There is no rank four and five modular tensor category enlarging the rank three surviving modular tensor category. At rank six, there is one consistent  modular tensor category enlarging the surviving modular tensor category, the one with $SU(2)_3/\mbb Z_2\times SU(2)_2$ realization. Since this scenario is a Deligne product, we can easily compute the effective central charge. If the emergent Fibonacci line $W$ has quantum dimension $d_W=\frac{1+\sqrt5}2$, the allowed topological twists are $\theta_W=e^{\pm\frac{4\pi i}5}$. Thus, depending on the sign of the global dimension $D_\text{Fibonacci}=\pm\sqrt{\frac{5+\sqrt5}2}$, we get central charges
\begin{table}[H]
\begin{center}
\begin{tabular}{c|c|c}
$D_\text{Fibonacci}\backslash\theta_W$&$e^{+\frac{4\pi i}5}$&$e^{-\frac{4\pi i}5}$\\\hline
$+\sqrt{\frac{5+\sqrt5}2}$&$\frac{14}5$&$-\frac{14}5$\\
$-\sqrt{\frac{5+\sqrt5}2}$&$-\frac65$&$\frac65$
\end{tabular}.
\end{center}
\caption{Central charges (mod $8$) of the Fibonacci modular tensor category with $d_W=\frac{1+\sqrt5}2$}
\end{table}
If the emergent Fibonacci line has $d_W=\frac{1-\sqrt5}2$, we have central charges
\begin{table}[H]
\begin{center}
\begin{tabular}{c|c|c}
$D_\text{Fibonacci}\backslash\theta_W$&$e^{+\frac{2\pi i}5}$&$e^{-\frac{2\pi i}5}$\\\hline
$+\sqrt{\frac{5-\sqrt5}2}$&$\frac25$&$-\frac25$\\
$-\sqrt{\frac{5-\sqrt5}2}$&$-\frac{18}5$&$\frac{18}5$
\end{tabular}.
\end{center}
\caption{Central charges (mod $8$) of the Fibonacci modular tensor category with $d_W=\frac{1-\sqrt5}2$}
\end{table}

On the other hand, surviving lines give an independent contribution. The double braiding relation (and ``monotonicity'') predicts the surviving lines $\eta\to j,N\to k$ are associated to primaries with conformal dimensions
\[ h_j^\text{IR}=\frac92-l,\quad h_k^\text{IR}=\frac{23}{16}-\frac m2, \]
where $l,m\in\mbb N$. These give $T$-matrices (in the basis $\{1,j,k\}$)
\[ T=\begin{pmatrix}1&0&0\\0&-1&0\\0&0&e^{\frac{7\pi i}8}\end{pmatrix} \]
for even $m$, or
\[ T=\begin{pmatrix}1&0&0\\0&-1&0\\0&0&e^{-\frac{\pi i}8}\end{pmatrix} \]
for odd $m$ together with the topological $S$-matrix
\[ S=\pm\frac12\begin{pmatrix}1&1&\sqrt2\\1&1&-\sqrt2\\\sqrt2&-\sqrt2&0\end{pmatrix}. \]
Thus, the rank three modular tensor category gives central charges (mod $8$)
\[ c^{\text{even }m}_+=\frac72,\quad c^{\text{even }m}_-=-\frac12,\quad c^{\text{odd }m}_+=-\frac12,\quad c^{\text{odd }m}_-=\frac72 \]
for positive and negative global dimensions.

In total, there are $2\times4\times4=32$ cases. Here we only consider one case corresponding to the bosonic $(5,4)$ minimal model, and see it is consistent with the effective $c$-theorem. This case corresponds to
\[ d_W=\frac{1+\sqrt5}2,\quad D_\text{Fibonacci}=+\sqrt{\frac{5+\sqrt5}2},\quad\theta_W=e^{-\frac{4\pi i}5},\quad m=2M. \]
In this case, the central charge is given by
\[ c=-\frac{14}5+\frac72=\frac7{10}\quad(\text{mod }8). \]
From the upper bound $c\le\frac{39}{10}$ in \cite{GK94}, we in general have $c=\frac7{10}-8n$ with $n\in\mbb N$. The conformal dimensions of primaries are given by
\[ h_1=0,\quad h_j=\frac92-l,\quad h_k=\frac{23}{16}-M,\quad h_W=-\frac25+p,\quad h_{jW}=\frac1{10}+q,\quad h_{kW}=\frac3{80}+r, \]
where $p,q,r\in\mbb Z$. Thus, the effective central charge is given by
\begin{align*}
    c^\text{eff}&=\left(\frac7{10}-8n\right)-24\min\left(0,\frac92-l,\frac{23}{16}-M,-\frac25+p,\frac1{10}+q,\frac3{80}+r\right)\\
    &=\frac1{10}\Big\{\left(7-80n\right)-3\min\left(0,360-80l,115-80M,-16+80p,8+80q,3+80r\right)\Big\}.
\end{align*}
The effective $c$-theorem $c^\text{eff}\le c^\text{eff}_\text{UV}=\frac{19}{22}\approx0.86$ imposes $b:=\{\}=0,1,\dots,8$. This has a solution (say $n=0$ and $h_{\min}=0$), and the rank six scenario is consistent with the effective $c$-theorem.

\underline{Spin contents}
\begin{align*}
    \mcal H_\eta:&\quad s\in\{0,\pm\frac12,\pm\frac32,\pm\frac52,\pm\frac72,\pm\frac92\}=\{0,\pm\frac12\}\text{ mod }1,\\
    \mcal H_N:&\quad s\in\{\pm\frac1{16},\pm\frac7{16},\pm\frac9{16},\pm\frac{15}{16},\pm\frac{17}{16},\pm\frac{23}{16},\pm\frac{25}{16},\pm\frac{31}{16},\pm\frac{39}{16},\pm\frac{47}{16}\}=\{\pm\frac1{16},\pm\frac7{16}\}\text{ mod }1.
\end{align*}
Spin constraint is also satisfied by the scenario that the bosonic $(11,4)$ minimal model flow to the  bosonic $(5,4)$ minimal model. With fermionization, this analysis gives further support to our fermionic renormalization group flow discussed in the main text.

\subsection{Bosonic $(E_6,A_6)$ minimal model}
It is the E-series modular invariant $(12,7)$ minimal model.
It has $c=-\frac{11}{14},c_\text{eff}=\frac{13}{14}\approx0.928$.

\underline{Primary operator and topological defect lines}
\begin{align*}
    C_{1,1}:=\chi_{1,1}+\chi_{1,7}\leftrightarrow\ep_0&\leftrightarrow1,\\
    C_{2,1}:=\chi_{2,1}+\chi_{2,7}\leftrightarrow\ep_{\frac{11}{14}}&\leftrightarrow\mcal L_2,\\
    C_{3,1}:=\chi_{3,1}+\chi_{3,7}\leftrightarrow\ep_{\frac37}&\leftrightarrow\mcal L_3,\\
    C_{1,4}:=\chi_{1,4}+\chi_{1,8}\leftrightarrow\sigma_{\frac{11}{16}}&\leftrightarrow N,\\
    C_{2,4}:=\chi_{2,4}+\chi_{2,8}\leftrightarrow\sigma_{-\frac3{112}}&\leftrightarrow\mcal L_5=N\mcal L_2,\\
    C_{3,4}:=\chi_{3,4}+\chi_{3,8}\leftrightarrow\sigma_{\frac{13}{112}}&\leftrightarrow\mcal L_6=N\mcal L_3,\\
    C_{1,5}:=\chi_{1,5}+\chi_{1,11}\leftrightarrow\ep_{\frac32}&\leftrightarrow\eta,\\
    C_{2,5}:=\chi_{2,5}+\chi_{2,11}\leftrightarrow\ep_{\frac27}&\leftrightarrow\mcal L_8=\eta\mcal L_2,\\
    C_{3,5}:=\chi_{3,5}+\chi_{3,11}\leftrightarrow\ep_{-\frac1{14}}&\leftrightarrow\mcal L_9=\eta\mcal L_3.
\end{align*}

\underline{Action of topological defect lines on primaries}
\begin{table}[H]
\begin{center}
\begin{tabular}{c||c|c|c|c|c|c|c|c|c}
&$id$&$\ep_{\frac{11}{14}}$&$\ep_{\frac37}$&$\sigma_{\frac{11}{16}}$&$\sigma_{-\frac3{112}}$&$\sigma_{\frac{13}{112}}$&$\ep_{\frac32}$&$\ep_{\frac27}$&$\ep_{-\frac1{14}}$\\\hline
$\eta$&$1$&$1$&$1$&$-1$&$-1$&$-1$&$1$&$1$&$1$\\
$N$&$-\sqrt2$&$\sqrt2$&$-\sqrt2$&$0$&$0$&$0$&$\sqrt2$&$-\sqrt2$&$\sqrt2$\\
$\mcal L_2$&$-2\sin\frac{3\pi}{14}$&$2\sin\frac\pi{14}$&$\frac{\cos\frac{3\pi}{14}}{\sin\frac\pi7}$&$2\sin\frac{3\pi}{14}$&$-2\sin\frac\pi{14}$&$-2\cos\frac\pi7$&$-2\sin\frac{3\pi}{14}$&$2\sin\frac\pi{14}$&$\frac{\cos\frac{3\pi}{14}}{\sin\frac\pi7}$\\
$\mcal L_3$&$\frac{\sin\frac\pi7}{\cos\frac{3\pi}{14}}$&$-\frac1{2\sin\frac{3\pi}{14}}$&$\frac1{2\sin\frac\pi{14}}$&$\frac{\sin\frac\pi7}{\cos\frac{3\pi}{14}}$&$-\frac1{2\sin\frac{3\pi}{14}}$&$\frac1{2\sin\frac\pi{14}}$&$\frac{\sin\frac\pi7}{\cos\frac{3\pi}{14}}$&$-\frac1{2\sin\frac{3\pi}{14}}$&$\frac1{2\sin\frac\pi{14}}$
\end{tabular}.
\end{center}
\end{table}

\underline{Spin content}
\begin{align*}
    \mcal H_\eta:&\quad s\in\{0,\pm\frac12,\pm\frac32\}=\{0,\pm\frac12\}\text{ mod }1,\\
    \mcal H_N:&\quad s\in\{\pm\frac3{16},\pm\frac5{16},\pm\frac{11}{16},\pm\frac{13}{16}\}=\{\pm\frac3{16},\pm\frac5{16}\}\text{ mod }1,\\
    \mcal H_{\mcal L_2}:&\quad s\in\{0,\pm\frac17,\pm\frac5{14},\pm\frac12,\pm\frac57,\pm\frac{11}{14},\pm\frac{17}{14}\}=\{0,\pm\frac17,\pm\frac3{14},\pm\frac27,\pm\frac5{14},\pm\frac12\}\text{ mod }1,\\
    \mcal H_{\mcal L_3}:&\quad s\in\{0,\pm\frac17,\pm\frac37,\pm\frac47,\pm\frac67,\pm\frac{11}7\}=\{0,\pm\frac17,\pm\frac37\}\text{ mod }1.
\end{align*}
With these data, let us consider whether the ultraviolet theory can flow to the bosonic $(5,4)$ minimal model in the infrared. The ultraviolet theory has the Tambara-Yamagami category, but the quantum dimension of $N$ does not match that of the bosonic $(5,4)$ minimal model. We also realize that the spin content of $N$ does not match. Thus we cannot realize a flow to the $(5,4)$ minimal model without breaking the $N$ line and hence violating our assumptions.

\subsection{Bosonic $(E_6,A_{10})$ minimal model}
It is the E-series modular invariant $(12,11)$ minimal model with  $c=\frac{21}{22}\approx0.955$.  It is unitary.

\underline{Primary operators and Verlinde lines}\newline
The theory has 15 primaries
\begin{align*}
    C_{1,1}:= \chi_{1,1}+\chi_{1,7}\leftrightarrow\ep_0&\leftrightarrow1,\\
    C_{3,1}:=\chi_{3,1}+\chi_{3,7}\leftrightarrow\ep_{\frac{13}{11}}&\leftrightarrow\mcal L_2,\\
    C_{5,1}:=\chi_{5,1}+\chi_{5,7}\leftrightarrow\ep_{\frac6{11}}&\leftrightarrow\mcal L_3,\\
   {C_{4,5}}: = \chi_{7,1}+\chi_{7,7}\leftrightarrow\ep_{\frac1{11}}&\leftrightarrow\mcal L_4,\\
    {C_{2,5}}:= \chi_{9,1}+\chi_{9,7}\leftrightarrow\ep_{\frac{20}{11}}&\leftrightarrow\mcal L_5,\\
    C_{1,4}:=\chi_{1,4}+\chi_{1,8}\leftrightarrow\sigma_{\frac{31}{16}}&\leftrightarrow N,\\
    C_{3,4}:=\chi_{3,4}+\chi_{3,8}\leftrightarrow\sigma_{\frac{21}{176}}&\leftrightarrow N\mcal L_2,\\
    C_{5,4}:=\chi_{5,4}+\chi_{5,8}\leftrightarrow\sigma_{\frac{85}{176}}&\leftrightarrow N\mcal L_3,\\
    {C_{4,4}} := \chi_{7,4}+\chi_{7,8}\leftrightarrow\sigma_{\frac5{176}}&\leftrightarrow N\mcal L_4,\\
    {C_{2,4}} := \chi_{9,4}+\chi_{9,8}\leftrightarrow\sigma_{\frac{133}{176}}&\leftrightarrow N\mcal L_5,\\
    C_{1,5}:=\chi_{1,5}+\chi_{1,11}\leftrightarrow\ep_{\frac72}&\leftrightarrow\eta,\\
    C_{3,5}:=\chi_{3,5}+\chi_{3,11}\leftrightarrow\ep_{\frac{15}{22}}&\leftrightarrow\eta\mcal L_2,\\
    C_{5,5}:=\chi_{5,5}+\chi_{5,11}\leftrightarrow\ep_{\frac1{22}}&\leftrightarrow\eta\mcal L_3,\\
    {C_{4,1}}: = \chi_{7,5}+\chi_{7,11}\leftrightarrow\ep_{\frac{35}{22}}&\leftrightarrow\eta\mcal L_4,\\
    {C_{2,1}}: = \chi_{9,5}+\chi_{9,11}\leftrightarrow\ep_{\frac7{22}}&\leftrightarrow\eta\mcal L_5.
\end{align*}

\underline{Actions of topological line defects on primaries}\newline
\begin{table}[H]
\hspace{-60pt}
\scalebox{0.69}{\begin{tabular}{c||c|c|c|c|c|c|c|c|c|c|c|c|c|c|c}
&$id$&$\ep_{\frac{13}{11}}$&$\ep_{\frac6{11}}$&$\ep_{\frac1{11}}$&$\ep_{\frac{20}{11}}$&$\sigma_{\frac{31}{16}}$&$\sigma_{\frac{21}{176}}$&$\sigma_{\frac{85}{176}}$&$\sigma_{\frac5{176}}$&$\sigma_{\frac{133}{176}}$&$\ep_{\frac72}$&$\ep_{\frac{15}{22}}$&$\ep_{\frac1{22}}$&$\ep_{\frac{35}{22}}$&$\ep_{\frac7{22}}$\\\hline
$\eta$&1&1&1&1&1&$-1$&$-1$&$-1$&$-1$&$-1$&$1$&$1$&$1$&$1$&$1$\\\hline
$N$&$\sqrt2$&$\sqrt2$&$\sqrt2$&$\sqrt2$&$\sqrt2$&$0$&$0$&$0$&$0$&$0$&$-\sqrt2$&$-\sqrt2$&$-\sqrt2$&$-\sqrt2$&$-\sqrt2$\\\hline
$\mcal L_2$&$\frac{\cos\frac{5\pi}{22}}{\sin\frac\pi{11}}$&$\frac{\sin\frac{2\pi}{11}}{\cos\frac{5\pi}{22}}$&$1-2\cos\frac\pi{11}$&$-\frac{\sin\frac\pi{11}}{\cos\frac{3\pi}{22}}$&$\frac{\cos\frac\pi{22}}{\sin\frac{2\pi}{11}}$&$\frac{\cos\frac{5\pi}{22}}{\sin\frac\pi{11}}$&$\frac{\sin\frac{2\pi}{11}}{\cos\frac{5\pi}{22}}$&$1-2\cos\frac\pi{11}$&$-\frac{\sin\frac\pi{11}}{\cos\frac{3\pi}{22}}$&$\frac{\cos\frac\pi{22}}{\sin\frac{2\pi}{11}}$&$\frac{\cos\frac{5\pi}{22}}{\sin\frac\pi{11}}$&$\frac{\sin\frac{2\pi}{11}}{\cos\frac{5\pi}{22}}$&$1-2\cos\frac\pi{11}$&$-\frac{\sin\frac\pi{11}}{\cos\frac{3\pi}{22}}$&$\frac{\cos\frac\pi{22}}{\sin\frac{2\pi}{11}}$\\\hline
$\mcal L_3$&$\frac1{2\sin\frac\pi{22}}$&$-\frac1{2\sin\frac{3\pi}{22}}$&$\frac1{2\sin\frac{5\pi}{22}}$&$-\frac{\sin\frac{2\pi}{11}}{\cos\frac{3\pi}{22}}$&$\frac1{2\cos\frac\pi{11}}$&$\frac1{2\sin\frac\pi{22}}$&$-\frac1{2\sin\frac{3\pi}{22}}$&$\frac1{2\sin\frac{5\pi}{22}}$&$-\frac{\sin\frac{2\pi}{11}}{\cos\frac{3\pi}{22}}$&$\frac1{2\cos\frac\pi{11}}$&$\frac1{2\sin\frac\pi{22}}$&$-\frac1{2\sin\frac{3\pi}{22}}$&$\frac1{2\sin\frac{5\pi}{22}}$&$-\frac{\sin\frac{2\pi}{11}}{\cos\frac{3\pi}{22}}$&$\frac1{2\cos\frac\pi{11}}$\\\hline
$\mcal L_4$&$\frac{\cos\frac{3\pi}{22}}{\sin\frac\pi{11}}$&$-\frac{\sin\frac\pi{11}}{\cos\frac{5\pi}{22}}$&$-\frac{\sin\frac{2\pi}{11}}{\cos\frac\pi{22}}$&$\frac1{-1+2\cos\frac\pi{11}}$&$-\frac{\cos\frac{5\pi}{22}}{\sin\frac{2\pi}{11}}$&$\frac{\cos\frac{3\pi}{22}}{\sin\frac\pi{11}}$&$-\frac{\sin\frac\pi{11}}{\cos\frac{5\pi}{22}}$&$-\frac{\sin\frac{2\pi}{11}}{\cos\frac\pi{22}}$&$\frac1{-1+2\cos\frac\pi{11}}$&$-\frac{\cos\frac{5\pi}{22}}{\sin\frac{2\pi}{11}}$&$\frac{\cos\frac{3\pi}{22}}{\sin\frac\pi{11}}$&$-\frac{\sin\frac\pi{11}}{\cos\frac{5\pi}{22}}$&$-\frac{\sin\frac{2\pi}{11}}{\cos\frac\pi{22}}$&$\frac1{-1+2\cos\frac\pi{11}}$&$-\frac{\cos\frac{5\pi}{22}}{\sin\frac{2\pi}{11}}$\\\hline
$\mcal L_5$&$2\cos\frac\pi{11}$&$2\sin\frac{5\pi}{22}$&$2\sin\frac\pi{22}$&$-2\sin\frac{3\pi}{22}$&$-\frac{\cos\frac{3\pi}{22}}{\sin\frac{2\pi}{11}}$&$2\cos\frac\pi{11}$&$2\sin\frac{5\pi}{22}$&$2\sin\frac\pi{22}$&$-2\sin\frac{3\pi}{22}$&$-\frac{\cos\frac{3\pi}{22}}{\sin\frac{2\pi}{11}}$&$2\cos\frac\pi{11}$&$2\sin\frac{5\pi}{22}$&$2\sin\frac\pi{22}$&$-2\sin\frac{3\pi}{22}$&$-\frac{\cos\frac{3\pi}{22}}{\sin\frac{2\pi}{11}}$
\end{tabular}.}
\end{table}

\underline{Spin content}
\begin{align*}
    \mcal H_\eta:&\quad s\in\{0,\pm\frac12,\pm\frac32,\pm\frac72\}=\{0,\pm\frac12\}\text{ mod }1,\\
    \mcal H_N:&\quad s\in\{\pm\frac1{16},\pm\frac7{16},\pm\frac9{16},\pm\frac{17}{16},\pm\frac{25}{16},\pm\frac{31}{16}\}=\{\pm\frac1{16},\pm\frac7{16}\}\text{ mod }1,\\
    \mcal H_{\mcal L_2}:&\quad s\in\{0,\pm\frac4{11},\pm\frac5{11},\pm\frac7{11},\pm\frac8{11},\pm\frac{13}{11},\pm\frac{14}{11},\pm\frac{17}{11},\pm\frac{19}{11},\pm\frac{20}{11},\pm\frac{31}{11}\}\\
    &\hspace{30pt}=\{0,\pm\frac2{11},\pm\frac3{11},\pm\frac4{11},\pm\frac5{11}\}\text{ mod }1,\\
    \mcal H_{\mcal L_3}:&\quad s\in\{0,\pm\frac1{11},\pm\frac3{11},\pm\frac4{11},\pm\frac5{11},\pm\frac6{11},\pm\frac7{11},\pm\frac8{11},\pm\frac{10}{11},\pm\frac{12}{11},\pm\frac{14}{11},\pm\frac{16}{11},\pm\frac{17}{11},\pm\frac{19}{11},\pm\frac{38}{11}\}\\
    &\hspace{30pt}=\{0,\pm\frac1{11},\pm\frac3{11}\pm\frac4{11},\pm\frac5{11}\}\text{ mod }1,\\
    \mcal H_{\mcal L_4}:&\quad s\in\{0,\pm\frac1{11},\pm\frac3{11},\pm\frac4{11},\pm\frac5{11},\pm\frac7{11},\pm\frac{10}{11},\pm\frac{12}{11},\pm\frac{14}{11},\pm\frac{17}{11},\pm\frac{21}{11}\}\\
    &\hspace{30pt}=\{0,\pm\frac1{11},\pm\frac3{11},\pm\frac4{11},\pm\frac5{11}\}\text{ mod }1,\\
    \mcal H_{\mcal L_5}:&\quad s\in\{0,\pm\frac1{11},\pm\frac4{11},\pm\frac5{11},\pm\frac7{11},\pm\frac{10}{11},\pm\frac{12}{11},\pm\frac{13}{11},\pm\frac{17}{11},\pm\frac{20}{11},\pm\frac{35}{11}\}\\
    &\hspace{30pt}=\{0,\pm\frac1{11},\pm\frac2{11},\pm\frac4{11},\pm\frac5{11}\}\text{ mod }1.
\end{align*}
{Let us see whether the ultraviolet theory can flow to the bosonic $(5,4)$ minimal model by the relevant deformation $\ep_{\frac6{11}},\ep_{\frac1{11}}$, which preserves the $N$ line. We observe that the $N$ line has the same quantum dimension as the one in the $N $ line of the bosonic $(5,4)$ minimal model. We also observe that the spin content is the same. At this point, we do not see any obstruction, but we have studied a possible obstruction from the double braiding relation in the main text (in section 5.5).}

\subsection{Bosonic $(A_{12},E_6)$ minimal model}

It is the E-series modular invariant $(13,12)$ minimal model with $c=\frac{25}{26}\approx0.96$. We will use the data below to study double braidings of surviving lines in the next subsection.

\underline{Primary operators and Verlinde lines}\newline
The theory has 18 primaries
\begin{align*}
    C_{1,1}:=\chi_{1,1}+\chi_{7,1}\leftrightarrow\ep_0&\leftrightarrow1,\\
    C_{1,3}:=\chi_{1,3}+\chi_{7,3}\leftrightarrow\ep_{\frac{11}{13}}&\leftrightarrow\mcal L_2,\\
    C_{1,5}:=\chi_{1,5}+\chi_{7,5}\leftrightarrow\ep_{\frac{20}{13}}&\leftrightarrow\mcal L_3,\\
    {C_{5,6}} := \chi_{1,7}+\chi_{7,7}\leftrightarrow\ep_{\frac1{13}}&\leftrightarrow\mcal L_4,\\
    {C_{5,4}} := \chi_{1,9}+\chi_{7,9}\leftrightarrow\ep_{\frac6{13}}&\leftrightarrow\mcal L_5,\\
    {C_{5,2}}:= \chi_{1,11}+\chi_{7,11}\leftrightarrow\ep_{\frac{35}{13}}&\leftrightarrow\mcal L_6,\\
    C_{4,1}:=\chi_{4,1}+\chi_{8,1}\leftrightarrow\sigma_{\frac{41}{16}}&\leftrightarrow\mcal L_7=N,\\
    C_{4,3}:=\chi_{4,3}+\chi_{8,3}\leftrightarrow\sigma_{\frac{85}{208}}&\leftrightarrow\mcal L_8=N\mcal L_2,\\
    C_{4,5}:=\chi_{4,5}+\chi_{8,5}\leftrightarrow\sigma_{\frac{21}{208}}&\leftrightarrow\mcal L_9=N\mcal L_3,\\
    {C_{4,6}}:= \chi_{4,7}+\chi_{8,7}\leftrightarrow\sigma_{\frac{133}{208}}&\leftrightarrow\mcal L_{10}=N\mcal L_4,\\
    {C_{4,4}} :=\chi_{4,9}+\chi_{8,9}\leftrightarrow\sigma_{\frac5{208}}&\leftrightarrow\mcal L_{11}=N\mcal L_5,\\
    {C_{4,2}} :=\chi_{4,11}+\chi_{8,11}\leftrightarrow\sigma_{\frac{261}{208}}&\leftrightarrow\mcal L_{12}=N\mcal L_6,\\
    C_{5,1}:=\chi_{5,1}+\chi_{11,1}\leftrightarrow\ep_{\frac92}&\leftrightarrow\mcal L_{13}=\eta,\\
    C_{5,3}:=\chi_{5,3}+\chi_{11,3}\leftrightarrow\ep_{\frac{35}{26}}&\leftrightarrow\mcal L_{14}=\eta\mcal L_2,\\
    C_{5,5}:=\chi_{5,5}+\chi_{11,5}\leftrightarrow\ep_{\frac1{26}}&\leftrightarrow\mcal L_{15}=\eta\mcal L_3,\\
    {C_{1,6}}: = \chi_{5,7}+\chi_{11,7}\leftrightarrow\ep_{\frac{15}{26}}&\leftrightarrow\mcal L_{16}=\eta\mcal L_4,\\
    {C_{1,4}} :=\chi_{5,9}+\chi_{11,9}\leftrightarrow\ep_{\frac{51}{26}}&\leftrightarrow\mcal L_{17}=\eta\mcal L_5,\\
    {C_{1,2}} :=\chi_{5,11}+\chi_{11,11}\leftrightarrow\ep_{\frac5{26}}&\leftrightarrow\mcal L_{18}=\eta\mcal L_6.
\end{align*}

\underline{Actions of topological defect lines on primaries}\newline
\begin{table}[H]
\hspace{-60pt}
\scalebox{0.55}{\begin{tabular}{c||c|c|c|c|c|c|c|c|c|c|c|c|c|c|c|c|c|c}
&$id$&$\ep_{\frac{11}{13}}$&$\ep_{\frac{20}{13}}$&$\ep_{\frac1{13}}$&$\ep_{\frac6{13}}$&$\ep_{\frac{35}{13}}$&$\sigma_{\frac{41}{16}}$&$\sigma_{\frac{85}{208}}$&$\sigma_{\frac{21}{208}}$&$\sigma_{\frac{133}{208}}$&$\sigma_{\frac5{208}}$&$\sigma_{\frac{261}{208}}$&$\ep_{\frac92}$&$\ep_{\frac{35}{26}}$&$\ep_{\frac1{26}}$&$\ep_{\frac{15}{26}}$&$\ep_{\frac{51}{26}}$&$\ep_{\frac5{26}}$\\\hline
$\eta$&1&1&1&1&1&1&$-1$&$-1$&$-1$&$-1$&$-1$&$-1$&$1$&$1$&$1$&$1$&$1$&$1$\\\hline
$N$&$\sqrt2$&$\sqrt2$&$\sqrt2$&$\sqrt2$&$\sqrt2$&$\sqrt2$&$0$&$0$&$0$&$0$&$0$&$0$&$-\sqrt2$&$-\sqrt2$&$-\sqrt2$&$-\sqrt2$&$-\sqrt2$&$-\sqrt2$\\\hline
$\mcal L_2$&$1+2\cos\frac{2\pi}{13}$&$\frac{\cos\frac{5\pi}{26}}{\sin\frac{3\pi}{13}}$&$-\frac{\sin\frac{2\pi}{13}}{\cos\frac{3\pi}{26}}$&$1-2\cos\frac\pi{13}$&$\frac{\sin\frac\pi{13}}{\cos\frac{5\pi}{26}}$&$\frac{\cos\frac\pi{26}}{\sin\frac{2\pi}{13}}$&$1+2\cos\frac{2\pi}{13}$&$\frac{\cos\frac{5\pi}{26}}{\sin\frac{3\pi}{13}}$&$-\frac{\sin\frac{2\pi}{13}}{\cos\frac{3\pi}{26}}$&$1-2\cos\frac\pi{13}$&$\frac{\sin\frac\pi{13}}{\cos\frac{5\pi}{26}}$&$\frac{\cos\frac\pi{26}}{\sin\frac{2\pi}{13}}$&$1+2\cos\frac{2\pi}{13}$&$\frac{\cos\frac{5\pi}{26}}{\sin\frac{3\pi}{13}}$&$-\frac{\sin\frac{2\pi}{13}}{\cos\frac{3\pi}{26}}$&$1-2\cos\frac\pi{13}$&$\frac{\sin\frac\pi{13}}{\cos\frac{5\pi}{26}}$&$\frac{\cos\frac\pi{26}}{\sin\frac{2\pi}{13}}$\\\hline
$\mcal L_3$&$\frac{\cos\frac{3\pi}{26}}{\sin\frac\pi{13}}$&$-\frac{\sin\frac{2\pi}{13}}{\sin\frac{3\pi}{13}}$&$-\frac{\sin\frac\pi{13}}{\cos\frac{3\pi}{26}}$&$\frac{\cos\frac{5\pi}{26}}{\cos\frac\pi{26}}$&$-\frac{\cos\frac\pi{26}}{\cos\frac{5\pi}{26}}$&$\frac{\sin\frac{3\pi}{13}}{\sin\frac{2\pi}{13}}$&$\frac{\cos\frac{3\pi}{26}}{\sin\frac\pi{13}}$&$-\frac{\sin\frac{2\pi}{13}}{\sin\frac{3\pi}{13}}$&$-\frac{\sin\frac\pi{13}}{\cos\frac{3\pi}{26}}$&$\frac{\cos\frac{5\pi}{26}}{\cos\frac\pi{26}}$&$-\frac{\cos\frac\pi{26}}{\cos\frac{5\pi}{26}}$&$\frac{\sin\frac{3\pi}{13}}{\sin\frac{2\pi}{13}}$&$\frac{\cos\frac{3\pi}{26}}{\sin\frac\pi{13}}$&$-\frac{\sin\frac{2\pi}{13}}{\sin\frac{3\pi}{13}}$&$-\frac{\sin\frac\pi{13}}{\cos\frac{3\pi}{26}}$&$\frac{\cos\frac{5\pi}{26}}{\cos\frac\pi{26}}$&$-\frac{\cos\frac\pi{26}}{\cos\frac{5\pi}{26}}$&$\frac{\sin\frac{3\pi}{13}}{\sin\frac{2\pi}{13}}$\\\hline
$\mcal L_4$&$\frac1{2\sin\frac\pi{26}}$&$-\frac1{2\sin\frac{3\pi}{26}}$&$\frac1{2\sin\frac{5\pi}{26}}$&$-\frac{\sin\frac{3\pi}{13}}{\cos\frac\pi{26}}$&$\frac{\sin\frac{2\pi}{13}}{\cos\frac{5\pi}{26}}$&$-\frac1{2\cos\frac\pi{13}}$&$\frac1{2\sin\frac\pi{26}}$&$-\frac1{2\sin\frac{3\pi}{26}}$&$\frac1{2\sin\frac{5\pi}{26}}$&$-\frac{\sin\frac{3\pi}{13}}{\cos\frac\pi{26}}$&$\frac{\sin\frac{2\pi}{13}}{\cos\frac{5\pi}{26}}$&$-\frac1{2\cos\frac\pi{13}}$&$\frac1{2\sin\frac\pi{26}}$&$-\frac1{2\sin\frac{3\pi}{26}}$&$\frac1{2\sin\frac{5\pi}{26}}$&$-\frac{\sin\frac{3\pi}{13}}{\cos\frac\pi{26}}$&$\frac{\sin\frac{2\pi}{13}}{\cos\frac{5\pi}{26}}$&$-\frac1{2\cos\frac\pi{13}}$\\\hline
$\mcal L_5$&$\frac{\cos\frac{5\pi}{26}}{\sin\frac\pi{13}}$&$\frac1{1+2\cos\frac{2\pi}{13}}$&$-\frac{\cos\frac\pi{26}}{\cos\frac{3\pi}{26}}$&$\frac{\sin\frac{2\pi}{13}}{\cos\frac\pi{26}}$&$\frac{\sin\frac{3\pi}{13}}{\cos\frac{5\pi}{26}}$&$-\frac{\cos\frac{3\pi}{26}}{\sin\frac{2\pi}{13}}$&$\frac{\cos\frac{5\pi}{26}}{\sin\frac\pi{13}}$&$\frac1{1+2\cos\frac{2\pi}{13}}$&$-\frac{\cos\frac\pi{26}}{\cos\frac{3\pi}{26}}$&$\frac{\sin\frac{2\pi}{13}}{\cos\frac\pi{26}}$&$\frac{\sin\frac{3\pi}{13}}{\cos\frac{5\pi}{26}}$&$-\frac{\cos\frac{3\pi}{26}}{\sin\frac{2\pi}{13}}$&$\frac{\cos\frac{5\pi}{26}}{\sin\frac\pi{13}}$&$\frac1{1+2\cos\frac{2\pi}{13}}$&$-\frac{\cos\frac\pi{26}}{\cos\frac{3\pi}{26}}$&$\frac{\sin\frac{2\pi}{13}}{\cos\frac\pi{26}}$&$\frac{\sin\frac{3\pi}{13}}{\cos\frac{5\pi}{26}}$&$-\frac{\cos\frac{3\pi}{26}}{\sin\frac{2\pi}{13}}$\\\hline
$\mcal L_6$&$2\cos\frac\pi{13}$&$\frac{\cos\frac\pi{26}}{\sin\frac{3\pi}{13}}$&$2\sin\frac{3\pi}{26}$&$-2\sin\frac\pi{26}$&$-2\sin\frac{5\pi}{26}$&$-\frac{\cos\frac{5\pi}{26}}{\sin\frac{2\pi}{13}}$&$2\cos\frac\pi{13}$&$\frac{\cos\frac\pi{26}}{\sin\frac{3\pi}{13}}$&$2\sin\frac{3\pi}{26}$&$-2\sin\frac\pi{26}$&$-2\sin\frac{5\pi}{26}$&$-\frac{\cos\frac{5\pi}{26}}{\sin\frac{2\pi}{13}}$&$2\cos\frac\pi{13}$&$\frac{\cos\frac\pi{26}}{\sin\frac{3\pi}{13}}$&$2\sin\frac{3\pi}{26}$&$-2\sin\frac\pi{26}$&$-2\sin\frac{5\pi}{26}$&$-\frac{\cos\frac{5\pi}{26}}{\sin\frac{2\pi}{13}}$
\end{tabular}.}
\end{table}
We find the least relevant operator $|\ep_{\frac{11}{13}}|^2$ preserves three topological defect lines $\{1,\eta,N\}$, i.e., the Tambara-Yamagami category.

\underline{Spin content}
\begin{align*}
    \mcal H_\eta:&\quad s\in\{0,\pm\frac12,\pm\frac32,\pm\frac52,\pm\frac92\}=\{0,\pm\frac12\}\text{ mod }1,\\
    \mcal H_N:&\quad s\in\{\pm\frac1{16},\pm\frac7{16},\pm\frac9{16},\pm\frac{15}{16},\pm\frac{17}{16},\pm\frac{23}{16},\pm\frac{31}{16},\pm\frac{41}{16}\}=\{\pm\frac1{16},\pm\frac7{16}\}\text{ mod }1,\\
    \mcal H_{\mcal L_2}:&\quad s\in\{0,\pm\frac4{13},\pm\frac5{13},\pm\frac7{13},\pm\frac8{13},\pm\frac9{13},\pm\frac{11}{13},\pm\frac{16}{13},\pm\frac{17}{13},\pm\frac{18}{13},\pm\frac{19}{13},\pm\frac{23}{13},\pm\frac{28}{13},\pm\frac{29}{13},\pm\frac{41}{13}\}\\
    &\hspace{30pt}=\{0,\pm\frac2{13},\pm\frac3{13},\pm\frac4{13},\pm\frac5{13},\pm\frac6{13}\}\text{ mod }1,\\
    \mcal H_{\mcal L_3}:&\quad s\in\{0,\pm\frac1{13},\pm\frac3{13},\pm\frac4{13},\pm\frac5{13},\pm\frac7{13},\pm\frac8{13},\pm\frac9{13},\pm\frac{10}{13},\\
    &\hspace{30pt}\pm\frac{14}{13},\pm\frac{16}{13},\pm\frac{17}{13},\pm\frac{18}{13},\pm\frac{19}{13},\pm\frac{20}{13},\pm\frac{23}{13},\pm\frac{25}{13},\pm\frac{29}{13},\pm\frac{32}{13},\pm\frac{34}{13},\pm\frac{58}{13}\}\\
    &\hspace{50pt}=\{0,\pm\frac1{13},\pm\frac3{13},\pm\frac4{13},\pm\frac5{13},\pm\frac6{13}\}\text{ mod }1,\\
    \mcal H_{\mcal L_4}:&\quad s\in\{0,\pm\frac1{13},\pm\frac2{13},\pm\frac3{13},\pm\frac4{13},\pm\frac5{13},\pm\frac7{13},\pm\frac8{13},\pm\frac9{13},\pm\frac{10}{13},\\
    &\hspace{30pt}\pm\frac{14}{13},\pm\frac{15}{13},\pm\frac{17}{13},\pm\frac{18}{13},\pm\frac{19}{13},\pm\frac{25}{13},\pm\frac{34}{13},\pm\frac{51}{13}\}\\
    &\hspace{50pt}=\{0,\pm\frac1{13},\pm\frac2{13},\pm\frac3{13},\pm\frac4{13},\pm\frac5{13},\pm\frac6{13}\}\text{ mod }1,\\
    \mcal H_{\mcal L_5}:&\quad s\in\{0,\pm\frac1{13},\pm\frac2{13},\pm\frac3{13},\pm\frac5{13},\pm\frac6{13},\pm\frac7{13},\pm\frac8{13},\pm\frac{10}{13},\pm\frac{11}{13},\\
    &\hspace{30pt}\pm\frac{14}{13},\pm\frac{15}{13},\pm\frac{18}{13},\pm\frac{19}{13},\pm\frac{24}{13},\pm\frac{25}{13},\pm\frac{33}{13}\}\\
    &\hspace{50pt}=\{0,\pm\frac1{13},\pm\frac2{13},\pm\frac3{13},\pm\frac5{13},\pm\frac6{13}\}\text{ mod }1,\\
    \mcal H_{\mcal L_6}:&\quad s\in\{0,\pm\frac1{13},\pm\frac5{13},\pm\frac7{13},\pm\frac8{13},\pm\frac{11}{13},\pm\frac{14}{13},\pm\frac{15}{13},\pm\frac{17}{13},\pm\frac{19}{13},\pm\frac{24}{13},\pm\frac{25}{13},\pm\frac{35}{13},\pm\frac{56}{13}\}\\
    &\hspace{30pt}=\{0,\pm\frac1{13},\pm\frac2{13},\pm\frac4{13},\pm\frac5{13},\pm\frac6{13}\}\text{ mod }1.
\end{align*}

\subsection{Fermionic $(A_{12},E_6)$ minimal model}

This theory with  $c=\frac{25}{26}\approx0.96$ has three singlet relevant operators, so it cannot be a candidate for the Gross-Neveu-Yukawa fixed point, but the following analysis is useful to understand the renormalization group flow of the fermionic $(E_6,A_{10})$ model.

\underline{Primary operator and topological line defects}
\begin{align*}
    C_{1,1}\bar C_{1,1}=id_{0,0}&\leftrightarrow1,\\
    C_{1,3}\bar C_{1,3}=\ep_{\frac{11}{13}}\bar\ep_{\frac{11}{13}}&\leftrightarrow\mcal L_{\frac{11}{13},\frac{11}{13}},\\
    C_{1,5}\bar C_{1,5}=\ep_{\frac{20}{13}}\bar\ep_{\frac{20}{13}}&\leftrightarrow\mcal L_{\frac{20}{13},\frac{20}{13}},\\
    {C_{5,6}\bar{C}_{5,6}} = \ep_{\frac1{13}}\bar\ep_{\frac1{13}}&\leftrightarrow\mcal L_{\frac1{13},\frac1{13}},\\
    {C_{5,4}\bar{C}_{5,4}} = \ep_{\frac6{13}}\bar\ep_{\frac6{13}}&\leftrightarrow\mcal L_{\frac6{13},\frac6{13}},\\
    {C_{5,2}\bar{C}_{5,2}}=\ep_{\frac{35}{13}}\bar\ep_{\frac{35}{13}}&\leftrightarrow\mcal L_{\frac{35}{13},\frac{35}{13}},\\
    C_{5,1}\bar C_{5,1}=\ep_{\frac92}\bar\ep_{\frac92}&\leftrightarrow(-1)^F,\\
    C_{5,3}\bar C_{5,3}=\ep_{\frac{35}{26}}\bar\ep_{\frac{35}{26}}&\leftrightarrow\mcal L_{\frac{35}{26},\frac{35}{26}},\\
    C_{5,5}\bar C_{5,5}=\ep_{\frac1{26}}\bar\ep_{\frac1{26}}&\leftrightarrow\mcal L_{\frac1{26},\frac1{26}},\\
    {C_{1,6}\bar{C}_{1,6}} = \ep_{\frac{15}{26}}\bar\ep_{\frac{15}{26}}&\leftrightarrow\mcal L_{\frac{15}{26},\frac{15}{26}},\\
   {C_{1,4}\bar{C}_{1,4}} = \ep_{\frac{51}{26}}\bar\ep_{\frac{51}{26}}&\leftrightarrow\mcal L_{\frac{51}{26},\frac{51}{26}},\\
    {C_{1,2}\bar{C}_{1,2}} = \ep_{\frac5{26}}\bar\ep_{\frac5{26}}&\leftrightarrow\mcal L_{\frac5{26},\frac5{26}},\\
    C_{1,1}\bar C_{5,1}=\bar\ep_{\frac92}&,\\
    C_{5,1}\bar C_{1,1}=\ep_{\frac92}&,\\
    C_{1,3}\bar C_{5,3}=\ep_{\frac{11}{13}}\bar\ep_{\frac{35}{26}}&,\\
    C_{5,3}\bar C_{1,3}=\ep_{\frac{35}{26}}\bar\ep_{\frac{11}{13}}&,\\
    C_{1,5}\bar C_{5,5}=\ep_{\frac{20}{13}}\bar\ep_{\frac1{26}}&,\\
    C_{5,5}\bar C_{1,5}=\ep_{\frac1{26}}\bar\ep_{\frac{20}{13}}&,\\
   {C_{5,6}\bar{C}_{1,6}} = \ep_{\frac1{13}}\bar\ep_{\frac{15}{26}}&,\\
    {C_{1,6}\bar{C}_{5,6}} = \ep_{\frac{15}{26}}\bar\ep_{\frac1{13}}&,\\
    {C_{5,4}\bar{C}_{1,4}} = \ep_{\frac6{13}}\bar\ep_{\frac{51}{26}}&,\\
   {C_{1,4}\bar{C}_{5,4}} =\ep_{\frac{51}{26}}\bar\ep_{\frac6{13}}&,\\
   {C_{5,2}\bar{C}_{1,2}} =\ep_{\frac{35}{13}}\bar\ep_{\frac5{26}}&,\\
   {C_{1,2}\bar{C}_{5,2}} =\ep_{\frac5{26}}\bar\ep_{\frac{35}{13}}&.
\end{align*}

\underline{Actions of topological defect lines on primaries}\\
Solving the (modified) Cardy conditions, we find 24 topological defect lines (12 m-type and 12 q-type) with actions
\begin{table}[H]
\hspace{-60pt}
\scalebox{0.4}{\begin{tabular}{c||c|c|c|c|c|c|c|c|c|c|c|c||c|c|c|c|c|c|c|c|c|c|c|c}
	NS-NS&$id$&$|\ep_{\frac{11}{13}}|^2$&$|\ep_{\frac{20}{13}}|^2$&$|\ep_{\frac1{13}}|^2$&$|\ep_{\frac6{13}}|^2$&$|\ep_{\frac{35}{13}}|^2$&$|\ep_{\frac92}|^2$&$|\ep_{\frac{35}{26}}|^2$&$|\ep_{\frac1{26}}|^2$&$|\ep_{\frac{15}{26}}|^2$&$|\ep_{\frac{51}{26}}|^2$&$|\ep_{\frac5{26}}|^2$&$\bar \ep_{\frac92}$&$\ep_{\frac92}$&$\ep_{\frac{11}{13}}\bar\ep_{\frac{35}{26}}$&$\ep_{\frac{35}{26}}\bar\ep_{\frac{11}{13}}$&$\ep_{\frac{20}{13}}\bar\ep_{\frac1{26}}$&$\ep_{\frac1{26}}\bar\ep_{\frac{20}{13}}$&$\ep_{\frac1{13}}\bar\ep_{\frac{15}{26}}$&$\ep_{\frac{15}{26}}\bar\ep_{\frac1{13}}$&$\ep_{\frac6{13}}\bar\ep_{\frac{51}{26}}$&$\ep_{\frac{51}{26}}\bar\ep_{\frac6{13}}$&$\ep_{\frac{35}{13}}\bar\ep_{\frac5{26}}$&$\ep_{\frac{35}{13}}\bar\ep_{\frac5{26}}$\\\hline
	$(-1)^F$&$1$&$1$&$1$&$1$&$1$&$1$&$1$&$1$&$1$&$1$&$1$&$1$&$-1$&$-1$&$-1$&$-1$&$-1$&$-1$&$-1$&$-1$&$-1$&$-1$&$-1$&$-1$\\
	$\mcal L$&$1$&$1$&$1$&$1$&$1$&$1$&$-1$&$-1$&$-1$&$-1$&$-1$&$-1$&$1$&$-1$&$1$&$-1$&$1$&$-1$&$1$&$-1$&$1$&$-1$&$1$&$-1$\\
	$\mcal L_{\frac{11}{13},\frac{11}{13}}$&$1+2\cos\frac{2\pi}{13}$&$\frac{\cos\frac{5\pi}{26}}{\sin\frac{3\pi}{13}}$&$-\frac{\sin\frac{2\pi}{13}}{\cos\frac{3\pi}{26}}$&$1-2\cos\frac\pi{13}$&$\frac{\sin\frac\pi{13}}{\cos\frac{5\pi}{26}}$&$\frac{\cos\frac\pi{26}}{\sin\frac{2\pi}{13}}$&$1+2\cos\frac{2\pi}{13}$&$\frac{\cos\frac{5\pi}{26}}{\sin\frac{3\pi}{13}}$&$-\frac{\sin\frac{2\pi}{13}}{\cos\frac{3\pi}{26}}$&$1-2\cos\frac\pi{13}$&$\frac{\sin\frac\pi{13}}{\cos\frac{5\pi}{26}}$&$\frac{\cos\frac\pi{26}}{\sin\frac{2\pi}{13}}$&$1+2\cos\frac{2\pi}{13}$&$1+2\cos\frac{2\pi}{13}$&$\frac{\cos\frac{5\pi}{26}}{\sin\frac{3\pi}{13}}$&$\frac{\cos\frac{5\pi}{26}}{\sin\frac{3\pi}{13}}$&$-\frac{\sin\frac{2\pi}{13}}{\cos\frac{3\pi}{26}}$&$-\frac{\sin\frac{2\pi}{13}}{\cos\frac{3\pi}{26}}$&$1-2\cos\frac\pi{13}$&$1-2\cos\frac\pi{13}$&$\frac{\sin\frac\pi{13}}{\cos\frac{5\pi}{26}}$&$\frac{\sin\frac\pi{13}}{\cos\frac{5\pi}{26}}$&$\frac{\cos\frac\pi{26}}{\sin\frac{2\pi}{13}}$&$\frac{\cos\frac\pi{26}}{\sin\frac{2\pi}{13}}$\\
	$\mcal L_{\frac{20}{13},\frac{20}{13}}$&$\frac{\cos\frac{3\pi}{26}}{\sin\frac\pi{13}}$&$-\frac{\sin\frac{2\pi}{13}}{\sin\frac{3\pi}{13}}$&$-\frac{\sin\frac\pi{13}}{\cos\frac{3\pi}{26}}$&$\frac{\cos\frac{5\pi}{26}}{\cos\frac\pi{26}}$&$-\frac{\cos\frac\pi{26}}{\cos\frac{5\pi}{26}}$&$\frac{\sin\frac{3\pi}{13}}{\sin\frac{2\pi}{13}}$&$\frac{\cos\frac{3\pi}{26}}{\sin\frac\pi{13}}$&$-\frac{\sin\frac{2\pi}{13}}{\sin\frac{3\pi}{13}}$&$-\frac{\sin\frac\pi{13}}{\cos\frac{3\pi}{26}}$&$\frac{\cos\frac{5\pi}{26}}{\cos\frac\pi{26}}$&$-\frac{\cos\frac\pi{26}}{\cos\frac{5\pi}{26}}$&$\frac{\sin\frac{3\pi}{13}}{\sin\frac{2\pi}{13}}$&$\frac{\cos\frac{3\pi}{26}}{\sin\frac\pi{13}}$&$\frac{\cos\frac{3\pi}{26}}{\sin\frac\pi{13}}$&$-\frac{\sin\frac{2\pi}{13}}{\sin\frac{3\pi}{13}}$&$-\frac{\sin\frac{2\pi}{13}}{\sin\frac{3\pi}{13}}$&$-\frac{\sin\frac\pi{13}}{\cos\frac{3\pi}{26}}$&$-\frac{\sin\frac\pi{13}}{\cos\frac{3\pi}{26}}$&$\frac{\cos\frac{5\pi}{26}}{\cos\frac\pi{26}}$&$\frac{\cos\frac{5\pi}{26}}{\cos\frac\pi{26}}$&$-\frac{\cos\frac\pi{26}}{\cos\frac{5\pi}{26}}$&$-\frac{\cos\frac\pi{26}}{\cos\frac{5\pi}{26}}$&$\frac{\sin\frac{3\pi}{13}}{\sin\frac{2\pi}{13}}$&$\frac{\sin\frac{3\pi}{13}}{\sin\frac{2\pi}{13}}$\\
	$\mcal L_{\frac1{13},\frac1{13}}$&$\frac1{2\sin\frac\pi{26}}$&$-\frac1{2\sin\frac{3\pi}{26}}$&$\frac1{2\sin\frac{5\pi}{26}}$&$-\frac{\sin\frac{3\pi}{13}}{\cos\frac\pi{26}}$&$\frac{\sin\frac{2\pi}{13}}{\cos\frac{5\pi}{26}}$&$-\frac1{2\cos\frac\pi{13}}$&$\frac1{2\sin\frac\pi{26}}$&$-\frac1{2\sin\frac{3\pi}{26}}$&$\frac1{2\sin\frac{5\pi}{26}}$&$-\frac{\sin\frac{3\pi}{13}}{\cos\frac\pi{26}}$&$\frac{\sin\frac{2\pi}{13}}{\cos\frac{5\pi}{26}}$&$-\frac1{2\cos\frac\pi{13}}$&$\frac1{2\sin\frac\pi{26}}$&$\frac1{2\sin\frac\pi{26}}$&$-\frac1{2\sin\frac{3\pi}{26}}$&$-\frac1{2\sin\frac{3\pi}{26}}$&$\frac1{2\sin\frac{5\pi}{26}}$&$\frac1{2\sin\frac{5\pi}{26}}$&$-\frac{\sin\frac{3\pi}{13}}{\cos\frac\pi{26}}$&$-\frac{\sin\frac{3\pi}{13}}{\cos\frac\pi{26}}$&$\frac{\sin\frac{2\pi}{13}}{\cos\frac{5\pi}{26}}$&$\frac{\sin\frac{2\pi}{13}}{\cos\frac{5\pi}{26}}$&$-\frac1{2\cos\frac\pi{13}}$&$-\frac1{2\cos\frac\pi{13}}$\\
	$\mcal L_{\frac6{13},\frac6{13}}$&$\frac{\cos\frac{5\pi}{26}}{\sin\frac\pi{13}}$&$\frac1{1+2\cos\frac{2\pi}{13}}$&$-\frac{\cos\frac\pi{26}}{\cos\frac{3\pi}{26}}$&$\frac{\sin\frac{2\pi}{13}}{\cos\frac\pi{26}}$&$\frac{\sin\frac{3\pi}{13}}{\cos\frac{5\pi}{26}}$&$-\frac{\cos\frac{3\pi}{26}}{\sin\frac{2\pi}{13}}$&$\frac{\cos\frac{5\pi}{26}}{\sin\frac\pi{13}}$&$\frac1{1+2\cos\frac{2\pi}{13}}$&$-\frac{\cos\frac\pi{26}}{\cos\frac{3\pi}{26}}$&$\frac{\sin\frac{2\pi}{13}}{\cos\frac\pi{26}}$&$\frac{\sin\frac{3\pi}{13}}{\cos\frac{5\pi}{26}}$&$-\frac{\cos\frac{3\pi}{26}}{\sin\frac{2\pi}{13}}$&$\frac{\cos\frac{5\pi}{26}}{\sin\frac\pi{13}}$&$\frac{\cos\frac{5\pi}{26}}{\sin\frac\pi{13}}$&$\frac1{1+2\cos\frac{2\pi}{13}}$&$\frac1{1+2\cos\frac{2\pi}{13}}$&$-\frac{\cos\frac\pi{26}}{\cos\frac{3\pi}{26}}$&$-\frac{\cos\frac\pi{26}}{\cos\frac{3\pi}{26}}$&$\frac{\sin\frac{2\pi}{13}}{\cos\frac\pi{26}}$&$\frac{\sin\frac{2\pi}{13}}{\cos\frac\pi{26}}$&$\frac{\sin\frac{3\pi}{13}}{\cos\frac{5\pi}{26}}$&$\frac{\sin\frac{3\pi}{13}}{\cos\frac{5\pi}{26}}$&$-\frac{\cos\frac{3\pi}{26}}{\sin\frac{2\pi}{13}}$&$-\frac{\cos\frac{3\pi}{26}}{\sin\frac{2\pi}{13}}$\\
	$\mcal L_{\frac{35}{13},\frac{35}{13}}$&$2\cos\frac\pi{13}$&$\frac{\cos\frac\pi{26}}{\sin\frac{3\pi}{13}}$&$2\sin\frac{3\pi}{26}$&$-2\sin\frac\pi{26}$&$-2\sin\frac{5\pi}{26}$&$-\frac{\cos\frac{5\pi}{26}}{\sin\frac{2\pi}{13}}$&$2\cos\frac\pi{13}$&$\frac{\cos\frac\pi{26}}{\sin\frac{3\pi}{13}}$&$2\sin\frac{3\pi}{26}$&$-2\sin\frac\pi{26}$&$-2\sin\frac{5\pi}{26}$&$-\frac{\cos\frac{5\pi}{26}}{\sin\frac{2\pi}{13}}$&$2\cos\frac\pi{13}$&$2\cos\frac\pi{13}$&$\frac{\cos\frac\pi{26}}{\sin\frac{3\pi}{13}}$&$\frac{\cos\frac\pi{26}}{\sin\frac{3\pi}{13}}$&$2\sin\frac{3\pi}{26}$&$2\sin\frac{3\pi}{26}$&$-2\sin\frac\pi{26}$&$-2\sin\frac\pi{26}$&$-2\sin\frac{5\pi}{26}$&$-2\sin\frac{5\pi}{26}$&$-\frac{\cos\frac{5\pi}{26}}{\sin\frac{2\pi}{13}}$&$-\frac{\cos\frac{5\pi}{26}}{\sin\frac{2\pi}{13}}$
\end{tabular}.}
\end{table}

\underline{Spin content}
\begin{align*}
    \mcal H_{(-1)^F}:&\quad s\in\{0\},\\
    \mcal H_{\mcal L_{\frac{11}{13},\frac{11}{13}}}:&\quad s\in\{0,\pm\frac1{26},\pm\frac3{26},\pm\frac5{26},\pm\frac7{26},\pm\frac5{13},\pm\frac12,\pm\frac7{13},\pm\frac9{13},\pm\frac{19}{26},\pm\frac{21}{26},\pm\frac{11}{13},\pm\frac{25}{26},\\
    &\hspace{30pt}\pm\frac{17}{13},\pm\frac{35}{26},\pm\frac{18}{13},\pm\frac{19}{13},\pm\frac32,\pm\frac{23}{13},\pm\frac{49}{26},\pm\frac{29}{13},\pm\frac52,\pm\frac{41}{13},\pm\frac{95}{26}\}\\
    &\hspace{50pt}=\{0,\pm\frac1{26},\pm\frac3{26},\pm\frac2{13},\pm\frac5{26},\pm\frac3{13},\pm\frac7{26},\pm\frac4{13},\pm\frac9{26},\pm\frac5{13},\pm\frac6{13},\pm\frac12\}\text{ mod }1,\\
    \mcal H_{\mcal L_{\frac{20}{13},\frac{20}{13}}}:&\quad s\in\{0,\pm\frac1{26},\pm\frac3{26},\pm\frac5{26},\pm\frac7{26},\pm\frac5{13},\pm\frac{11}{26},\pm\frac{11}{26},\pm\frac12,\pm\frac7{13},\pm\frac9{13},\pm\frac{19}{26},\pm\frac{10}{13},\pm\frac{21}{26},\pm\frac{25}{26},\\
    &\hspace{30pt}\pm\frac{14}{13},\pm\frac{33}{26},\pm\frac{17}{13},\pm\frac{18}{13},\pm\frac{19}{13},\pm\frac32,\pm\frac{20}{13},\pm\frac{23}{13},\pm\frac{49}{26},\pm\frac{25}{13},\pm\frac{55}{26},\pm\frac{29}{13},\pm\frac{34}{13},\pm\frac{77}{26},\pm\frac{58}{13}\}\\
    &\hspace{50pt}=\{0,\pm\frac1{26},\pm\frac1{13},\pm\frac3{26},\pm\frac5{26},\pm\frac3{13},\pm\frac7{26},\pm\frac4{13},\pm\frac5{13},\pm\frac{11}{26},\pm\frac6{13},\pm\frac12\}\text{ mod }1,\\
    \mcal H_{\mcal L_{\frac1{13},\frac1{13}}}:&\quad s\in\{0,\pm\frac1{26},\pm\frac1{13},\pm\frac3{26},\pm\frac2{13},\pm\frac5{26},\pm\frac7{26},\pm\frac5{13},\pm\frac{11}{26},\pm\frac12,\pm\frac7{13},\pm\frac{15}{26},\pm\frac8{13},\pm\frac9{13},\pm\frac{10}{13},\pm\frac{21}{26},\pm\frac{23}{26},\pm\frac{25}{26},\\
    &\hspace{30pt}\pm\frac{14}{13},\pm\frac{29}{26},\pm\frac{15}{13},\pm\frac{33}{26},\pm\frac{17}{13},\pm\frac{35}{26},\pm\frac{18}{13},\pm\frac{19}{13},\pm\frac32,\pm\frac{49}{26},\pm\frac{25}{13},\pm\frac{55}{26},\pm\frac{34}{13},\pm\frac{69}{26},\pm\frac{51}{13},\pm\frac{115}{26}\}\\
    &\hspace{50pt}=\{0,\pm\frac1{26},\pm\frac1{13},\pm\frac3{26},\pm\frac2{13},\pm\frac5{26},\pm\frac3{13},\pm\frac7{26},\pm\frac4{13},\pm\frac9{26},\pm\frac5{13},\pm\frac{11}{26},\pm\frac6{13},\pm\frac12\}\text{ mod }1,\\
    \mcal H_{\mcal L_{\frac6{13},\frac6{13}}}:&\quad s\in\{0,\pm\frac1{26},\pm\frac3{26},\pm\frac2{13},\pm\frac7{26},\pm\frac5{13},\pm\frac{11}{26},\pm\frac6{13},\pm\frac12,\pm\frac7{13},\pm\frac8{13},\pm\frac{17}{26},\pm\frac{10}{13},\pm\frac{23}{26},\pm\frac{25}{26},\\
    &\hspace{30pt}\pm\frac{14}{13},\pm\frac{29}{26},\pm\frac{15}{13},\pm\frac{33}{26},\pm\frac{35}{26},\pm\frac{18}{13},\pm\frac{19}{13},\pm\frac32,\pm\frac{24}{13},\pm\frac{49}{26},\pm\frac{25}{13},\pm\frac{51}{26},\pm\frac{33}{13},\pm\frac{69}{26},\pm\frac{105}{26}\}\\
    &\hspace{50pt}=\{0,\pm\frac1{26},\pm\frac1{13},\pm\frac3{26},\pm\frac2{13},\pm\frac3{13},\pm\frac7{26},\pm\frac9{26},\pm\frac5{13},\pm\frac{11}{26},\pm\frac6{13},\pm\frac12\}\text{ mod }1,\\
    \mcal H_{\mcal L_{\frac{35}{13},\frac{35}{13}}}:&\quad s\in\{0,\pm\frac1{26},\pm\frac5{26},\pm\frac5{13},\pm\frac{11}{26},\pm\frac12,\pm\frac7{13},\pm\frac8{13},\pm\frac{17}{26},\pm\frac{23}{26},\pm\frac{25}{26},\\
    &\hspace{30pt}\pm\frac{14}{13},\pm\frac{29}{26},\pm\frac{15}{13},\pm\frac{35}{26},\pm\frac{19}{13},\pm\frac{47}{26},\pm\frac{24}{13},\pm\frac{25}{13},\pm\frac{35}{13},\pm\frac{56}{13}\}\\
    &\hspace{50pt}=\{0,\pm\frac1{26},\pm\frac1{13},\pm\frac3{26},\pm\frac2{13},\pm\frac5{26},\pm\frac4{13},\pm\frac9{26},\pm\frac5{13},\pm\frac{11}{26},\pm\frac6{13},\pm\frac12\}\text{ mod }1,\\
    \mcal H_{\mcal L}:&\quad s\in\{-\frac{31}{16},-\frac{23}{16},-\frac{35}{16},-\frac{15}{16},-\frac7{16},\frac1{16},\frac9{16},\frac{17}{16},\frac{41}{16}\}=\{-\frac7{16},\frac1{16}\}\text{ mod }1,\\
    \mcal H_{(-1)^F\mcal L}:&\quad s\in\{-\frac{41}{16},-\frac{17}{16},-\frac9{16},-\frac1{16},\frac7{16},\frac{15}{16},\frac{23}{16},\frac{31}{16}\}=\{-\frac1{16},\frac7{16}\}\text{ mod }1.
\end{align*}

{The relevant operators $|\ep_{\frac{11}{13}}|^2,|\ep_{\frac1{13}}|^2,|\ep_{\frac6{13}}|^2$ preserve the chiral $\mathbb{Z}_2$ symmetry. Consider the least relevant deformation by $|\ep_{\frac{11}{13}}|^2$. It is believed to be integrable and generate the renormalization group flow to the fermionic $(E_6,A_{10})$ minimal model. Under the renormalization group flow, the preserved topological defect line becomes
 \[
\begin{array}{ccccc}
    \text{fermionic }(A_{12},E_6):&1&(-1)^F&\mcal L&(-1)^F\mcal L\\
    &\downarrow&\downarrow&\downarrow&\downarrow\\
    \text{fermionic }(E_6,A_{10}):&1&(-1)^F&(-1)^F\mcal L&\mcal L
\end{array}.
\]
Indeed, one can check that the spin contents of preserved topological defect lines match. We also note that in the bosonic counterpart discussed in the previous subsection, the double braiding relation is satisfied:
{\begin{align*}
    c^{(A_{12},E_6)}_{N,N}c^{(A_{12},E_6)}_{N,N}&=e^{-\frac{\pi i}4}id_1\oplus e^{\frac{3\pi i}4}id_\eta,\\
    c^{(E_6,A_{10})}_{N,N}c^{(E_6,A_{10})}_{N,N}&=e^{\frac{\pi i}4}id_1\oplus e^{-\frac{3\pi i}4}id_\eta.
\end{align*}
This gives an example of double braiding relation beyond A-series minimal model, and gives another support of our discussion in section \ref{dbrconst}.}

\section{$\mathbb{Z}_2$ invariance of correlation functions on sphere from action of topological defect lines}\label{Z2}

In this appendix, we discuss the selection rule of correlation functions on a sphere from the (anomalous) $\mathbb{Z}_2$ symmetry from the viewpoint of the topological defect line.

Let us define the $\mathbb{Z}_2$ action of an invertible topological defect line (i.e. symmetry generating topological defect line) on states on the cylinder
\begin{align}
\hat{\mathcal{L}}|\phi_{\pm} \rangle = \pm |\phi_{\pm} \rangle \ .
\end{align}
We have already seen that in non-unitary theories, the invertible topological defect line may act on the vacuum with $(-1)$ rather than $+1$.
It induces the action on local operators on the sphere
\begin{align}
\mathcal{L}(\phi_{\pm}) = \pm (-1)^B \phi_{\pm} \ ,
\end{align}
where the presence of non-zero $B$ indicates the isotropy anomaly \cite{Chang:2018iay}. Let us also define the 't Hooft anomaly of the invertible topological defect line by the phase $\alpha = (-1)^A$ appearing Fig 6 of \cite{Lin:2019kpn}.

Now let us consider a correlation function $\langle \mathcal{L}(\phi_1,\cdots \phi_n) \rangle$ on the sphere. By shrinking $\mathcal{L}$ at the north pole, where we assume all operators reside in the southern hemisphere, we have
\begin{align}
\langle \mathcal{L}(\phi_1,\cdots \phi_n) \rangle = (-1)^B (-1)^{C}\langle \phi_1,\cdots \phi_n \rangle \ ,
\end{align}
where $\hat{\mathcal{L}}|1 \rangle = (-1)^{C}|1\rangle$. 

On the other hand, by repeatedly using the move of Fig 6 of \cite{Lin:2019kpn} and encircling each operator $\phi_i$ one by one, we obtain
\begin{align}
\langle \mathcal{L}(\phi_1,\cdots \phi_n) \rangle = (-1)^{A(n-1)}(-1)^{n_-} (-1)^{Bn}\langle \phi_1,\cdots \phi_n \rangle \ ,
\end{align}
where $n_\pm$ is the number of $\phi_{\pm}$'s so that $n=n_++n_-$.

By equating the two computations, we obtain the selection rule
\begin{align}
B+C \equiv A(n-1) + n_- + Bn
\end{align}
modulo 2. 

When the $\mathbb{Z}_2$ symmetry is non-anomalous, we have $A=B=C =0$ and the selection rule is $n_- \equiv 0$ as expected. When the symmetry is anomalous but preserves the vacuum on the cylinder, we have $A=1$ and $C=0$  (an example of which is the chiral $\mathbb{Z}_2$ symmetry of $(4,3)$ and $(5,4)$ fermionic minimal model). In \cite{Chang:2018iay}, it was argued that $B=1$ in this case, so we have the selection rule $n_- \equiv 0$ again. 

When the topological defect line does not preserve the vacuum, we might expect that we cannot obtain any interesting selection rule, but there is an exception. Suppose the symmetry has no 't Hooft anomaly and does not preserve the vacuum on the cylinder, we have $A=0$, and $C=1$. When $B=0$, the selection rule is $n_- \equiv 1$ and this does not give any constraint because we can always insert $1$ (note that $\hat{\mathcal{L}}|1\rangle=-|1\rangle $) to satisfy the selection rule. When $B=1$, the selection rule is $n_+ \equiv 0$ and now it becomes a non-trivial constraint on the correlation functions. Indeed, it effectively means that the correlation functions on the sphere satisfy the $\mathbb{Z}_2$ symmetry.

Similarly, consider the case when the topological defect line has an 't Hooft anomaly and does not preserve the vacuum on the cylinder. We have $A=1$, and $C=1$. When 
$B=1$, we have $n_- \equiv 1$ and this does not give any constraint, but when $B=0$, we have $n_+=0$ and this gives an effective $\mathbb{Z}_2$ symmetry on correlation functions (example of which is (odd, odd) minimal models and the chiral $\mathbb{Z}_2$ symmetry in certain non-unitary fermionic minimal models).

\end{document}